\pdfoutput=1 %arXiv admin added this line

\documentclass[graybox]{svmult}

\usepackage{graphicx}

\usepackage{amsmath}  
\usepackage[latin1]{inputenc}
\usepackage{natbib}
\usepackage{graphicx}
\usepackage{times}
\usepackage{mathptmx}       % selects Times Roman as basic font
\usepackage{helvet}         % selects Helvetica as sans-serif font
\usepackage{courier}        % selects Courier as typewriter font
\usepackage{type1cm}        % activate if the above 3 fonts are
\usepackage{makeidx}         % allows index generation
\usepackage{graphicx}        % standard LaTeX graphics tool
\usepackage{multicol}        % used for the two-column index
\usepackage[bottom]{footmisc}% places footnotes at page bottom

\makeindex             % used for the subject index
\bibpunct{[}{]}{,}{a}{}{;}

\begin{document}

\title*{Red Queen Coevolution on Fitness Landscapes}
\titlerunning{Coevolution on Fitness Landscapes} %for an abbreviated version of
\author{Ricard V. Sol\'e and Josep Sardany\'es}
\institute{Ricard V. Sol\'e \at $^1$ICREA-Complex Systems Lab, Universitat Pompeu Fabra, Parc de Recerca Biom\`edica de Barcelona (PRBB), Dr. Aiguader 88, 08003 Barcelona, Spain. $^2$Institut de Biologia Evolutiva (CSIC-Universitat Pompeu Fabra), Passeig Mar\'itim de la Barceloneta 37, 08003 Barcelona, Spain. $^3$Santa Fe Institute, 1399 Hyde Park Road, Santa Fe NM 87501, USA. \email{ricard.sole@upf.edu}
\and Josep Sardany\'es \at $^1$ICREA-Complex Systems Lab, Universitat Pompeu Fabra, Parc de Recerca Biom\`edica de Barcelona (PRBB), Dr. Aiguader 88, 08003 Barcelona, Spain. $^2$Institut de Biologia Evolutiva (CSIC-Universitat Pompeu Fabra), Passeig Mar\'itim de la Barceloneta 37, 08003 Barcelona, Spain.
 \email{josep.sardanes@upf.edu}}
\maketitle

\abstract{Species do not merely evolve, they also coevolve with other organisms. Coevolution is a major force driving interacting species to continuously evolve exploring their fitness landscapes. Coevolution involves the coupling of species fitness landscapes, linking species genetic changes with their inter-specific ecological interactions. Here we first introduce the Red Queen hypothesis of evolution commenting on some theoretical aspects and empirical evidences. As an introduction to the fitness landscape concept, we review key issues on evolution on simple and rugged fitness landscapes. Then we present key modeling examples of coevolution on different fitness landscapes at different scales, from RNA viruses to complex ecosystems and macroevolution.}

\section{Introduction: the Red Queen}
\label{sec:1}
Coevolution pervades evolutionary change on multiple scales. It is not exaggerated to say, 
changing a little the classical Dobzhansky's statement, that nothing makes sense 
in biology except in the light of coevolution. Darwin himself recognized this when referring 
to what he called the entangled bank \cite{Darwin1859}: "It is interesting to contemplate an entangled bank, clothed with many plants of many kinds, 
with birds singing on the bushes, with various insects flitting about, and with worms crawling through the damp earth". 
Indeed, ecosystems need to be seen as 
collectives of interacting species whose evolutionary fate is necessarily intermingled in complex ways. Such complex 
networks pervade ecosystems and their evolutionary dynamics \cite{Montoya2006}.

The rationale of the previous statements is simple. As any species changes over time, it inevitably 
triggers co-evolutionary responses  in those partners directly affected by their interdependencies. A prey running faster 
than its predator will require changes in the later to cope with the change. Running fast is one option, hiding 
in the appropriate place another. If we move ourselves into the microcosmos of host-pathogen interactions, 
including cells and viruses (or bacteria) as main examples of coevolving players, a faster rate of change in the parasite might need a faster response from the host. Changing attributes might be an unavoidable consequence 
of entangled ecosystems and not changing might possibly be lethal. This view matches a 
dynamically unstable scenario where changes keep happening all the time and, as 
in Lewis Carroll's {\em Through the looking glass}, species need -as Alice does- to {\em constantly run to remain in the same place}. 
Such picture was early supported by Leigh Van Valen's work, and is known
as the Red Queen hypothesis. 

The Red Queen model was introduced by Leigh Van Valen in 1973 \cite{vanValen1973}. 
It was conceived as a theoretical explanation for the observation that the extinction probability of a species is approximately 
independent of its length of existence \cite{vanValen1973,Benton1995}. Accordingly with this view, Van Valen observed that the vast majority of taxonomic groups analyzed displayed exponentially decaying survivorship curves. This result implied constancy in the probability of extinction of the taxa, regardless of their previous duration. That is, both data from the fossil record and from extant species suggested that a given species may disappear at any time, irrespective of how long has already existed. This unexpected phenomenon, termed the {\em Law of Constant Extinction}, can be formulated, in a simple way, as follows. If $N(t)$ indicates the number of species at a given time and we follow their presence over time (ignoring other events) we would observe an exponential decay law, namely: $$\frac{dN}{dt} = - \delta(t) N,$$
where $\delta(t)$ indicates a time-dependent extinction rate. If $N_0$ is the original number, this differential equation is easily solved, and gives: $$N(t) = N_0 \exp \left[-\int_0^t \delta(t) dt\right].$$ Despite the seemingly obvious assumption that $\delta$ depends on $t$, the surprising observation is that the observed curves fit very well a constant decay rate $\delta$, i.e., a solution: $$N(t)=N_0 e^{-\delta t},$$ 
where $\delta$ is the extinction probability of a species (per millions of years, Myr). 
\begin{figure*}
\center
\includegraphics[width=1.00\textwidth]{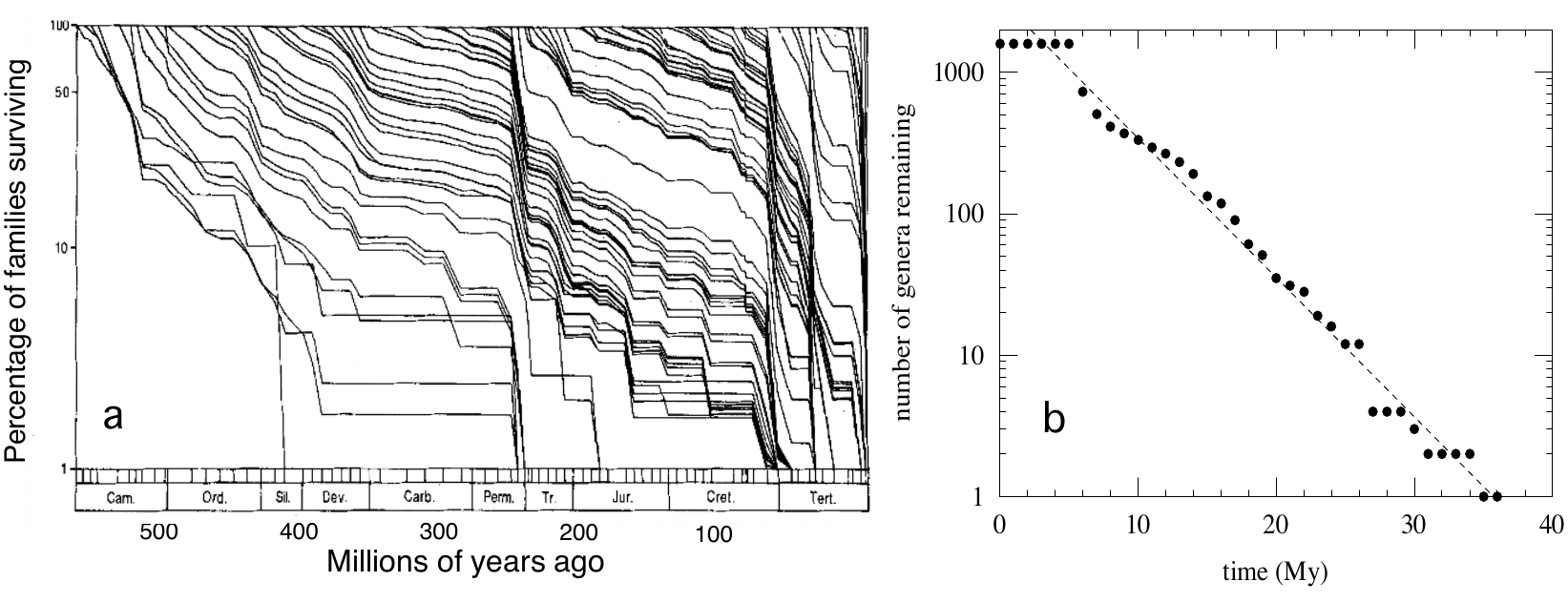}
\caption{Extinction and the fossil record of life. (a) Successive cohort survivorship curves for 
2,316 extinct marine families during the Phanerozoic (redrawn after \cite{Raup1991}). 
Notice that, together with an average exponential tendency to
decline (which would give straight lines in
a linear-log plot), discrete and punctuated events are found in the dynamics of survivorship. This pattern is well appreciated 
in (b) where we display the number of ammonoid genera surviving over geologic time. }
\label{figure1}
\end{figure*}
This law is essentially correct on average, despite the fine-scale pattern is much more episodic \cite{Raup1986}, as depicted in figure \ref{figure1}(a). Here we display the surviving sets of families found in the marine fossil record (the so called {\em seudocohorts}) through time. A roughly exponential decay can be identified, together with sharp extinctions events (see also figure \ref{figure1}(b)).

Our intuition, guided by Darwin's theory of natural selection, would have expected species within
any group to become longer lived along time: if adaptation improves
species progressively through time, a decreasing probability of extinction
should be expected. That is, older species should last longer. However, careful
analysis revealed that species of modern mammals 
are likely to become extinct as were their ancestors living
200 Myr ago \cite{Benton1995}. If evolution leads to improvement through adaptation,
why modern mammals have the same extinction probabilities as their ancestors?
Van Valen's interpretation is simple but counterintuitive: species do not evolve
to become any better at avoiding extinction. Van Valen suggested
that constant extinction probability would arise in an always changing 
biotic community, with species continually adapting to each other's changes. The name for his conjecture alludes to the Red Queen's
remark in Lewis Carroll's {\em Alice Through the Looking Glass}: 
``here, you see, it takes all the running you can do, to keep in the same place''. 
Van Valen's view of evolution is that species change just to remain in
the evolutionary game and extinction occurs when no further
changes are possible. Actually, the Red Queen hypothesis is profoundly Darwinian, in that it puts emphasis mostly on biotic interactions rather than on abiotic factors \cite{Hoffman1991}. Van Valen further elaborated this concept in much more detail in subsequent articles, showing that his theory was compatible with the classic population-genetic view of species evolution \cite{VanValen1976,VanValen1980}.

To test the plausibility of the Red Queen hypothesis, Maynard Smith and
co-workers (see \cite{Stenseth1984} and references therein) developed a theoretical model describing continuous (co-)evolution of species 
in a constant environment. Such model considered a fixed number of $S$ interacting species, defining some fitness measure $\phi$, and a maximum 
fitness $\phi_i^*$ was supposed to exist for each species in a given fixed, 
external biotic environment. At a given time, the fitness $\phi_i$ and 
the maximum $\phi_i^*$ took different values, and each species "tried"
to reduce the so called {\it lag load}, defined as:
\begin{equation}\label{lagload}
\nonumber{L_i = { \phi_i - \phi_i^* \over \phi_i } ; \; \; \; i=1,..., S.}
\end{equation}
If $\beta_{ij}$ is the change in the lag load $L_i$ due to a change
in $L_j$, then a mean-field equation for the average lag load 
$\langle L \rangle=\sum_i(L_i)/S$ can be derived. This is done by first 
separating, for each species, changes due to ``microevolution 
of coexisting species'' from those linked with 
its own microevolution \cite{Stenseth1984}. The
whole equation for the lag load variation in a given species is: $\delta L_i = \delta_c L_i - \delta_g L_i$, which simply says that it typically increases due to
changes in the other species and decreases due to microevolutionary
changes in the species under consideration i.e., 
$\delta_c L_i$ is the increase in the lag of the $i$th species caused by evolutionary changes in others, and $\delta_g L_i$ is the reduction in lag caused by changes in species $i$ itself. This can be written 
in the following way,
\begin{equation}\label{lagloadvar2}
\nonumber{
\delta L_i = \sum_{j=1}^S \beta_{ij} 
\delta_g L_j - \delta_g L_i, }
\end{equation}
where $\beta_{ij}$ (with $\beta_{ii}=0$) is the increase in $L_i$ due to
a (unit) change in $L_j$. Assuming that most species are close
to their adaptive peaks, any evolutionary change in one species
will have a deleterious effect on the other species. The deterministic, time-continuous equivalent model can be formulated with:
\begin{equation}
{d L_i \over dt} = \sum_{j=1}^S \beta_{ij} k_j L_j - k_i L_i.
\label{pre}
\end{equation}

By taking the average in both sides of Eq. \eqref{pre}, the evolution of
the average lag load is given by:
\begin{equation}\label{difavlag1}
\nonumber{
{d \langle L \rangle \over dt} = {1 \over S} \sum_{i=1}^S 
\Biggl \{ \sum_{j=1}^S \beta_{ij} k_j L_j - k_i L_i \Biggr \}}. 
\end{equation}
Assuming now that $k_i=k$ for all $i=1,...,S$, the
average lag load equation can be written as:
\begin{equation}\label{difavlag2}
\nonumber{
{d \langle L \rangle \over dt} = {k \over S} \sum_{j=1}^S ( \Psi_j - 1 ) L_j,}
\end{equation}
and it has a steady state solution if $\Psi_j=1$ for all $j=1,...,S$. In other
words, if:
\begin{equation}\label{Gamma}
\nonumber{
\Gamma \equiv  \sum_{i=1}^S \beta_{ij} = 1; 
\; \; \; \; \forall \; j. }
\end{equation}
Otherwise, it can be shown that $ \langle L \rangle $ will decrease (increase) 
for $\Gamma<1$ 
($\Gamma>1$). The previous identity is telling us that
the equilibrium state of this system is reached through a balance between
the reduction of the individual lag load of each species and the increases
due to coevolutionary changes in the remaining partners. The main result of this model is that evolution of species proceeds at an approximately steady
rate even in the absence of external or environmental changes \cite{Stenseth1984}. At the end of this chapter we will present a dynamic model of Red Queen dynamics 
where evolving networks interactions are made explicit.

As we previously discussed, the Red Queen hypothesis provides a plausible explanation of the fossil data record, but it turned to have more implications. For instance, one suggested implication of the Red Queen hypothesis is that coevolving pathogens may facilitate the persistence of outcrossing despite its costs. Specifically, coevolutionary interactions between hosts and pathogens might generate ever-changing conditions and thus favor the long-term maintenance of outcrossing relative to self-fertilization \cite{Agrawal2001} or asexual reproduction \cite{Jaenike1978,Hamilton1980} (see also \cite{Hamilton1990} for a review). Outcrossing (mating between different individuals) involves the introduction of unrelated genetic material into a breeding line, thus increasing genetic diversity. The previous statements are supported by evidences from nature. For instance, experimental studies on the coevolution of a nematode with a bacterial pathogen \cite{Morran2011} revealed that the action of parasites caused an increase of outcrossing in mixed mating populations. Interestingly, these experiments also revealed that coevolution with the pathogen caused extinction in populations without outcrossing, whereas outcrossing populations persisted through reciprocal coevolution. Studies in natural snails populations also revealed that sexual reproduction is more common when parasites are abundant and adapted to infect local host populations \cite{Lively1987,King2009}. Coevolution and Red Queen dynamics were also identified for the crustacean genus \emph{Daphnia} and its parasites in pond sediments \cite{Decaestecker2007}.

\section{Red Queen on a Lattice: a toy model}

Before we get into the more formal approaches taken to describe and simulate the evolution 
and coevolution of species on fitness landscapes, let us consider a simple toy model that illustrates 
the basic idea behind van Valen's metaphor. Imagine a world where our species can move on 
the surface of a sphere. To makes things simpler, consider a discretized surface, like a mesh covering the 
sphere\footnote{Specifically, we start from a lattice, whose surface has been discretized using a mesh, 
and then we perform a projection of this mesh on a surface by properly deforming the initial coordinates using a 
so called icosahedron-based pixelization. For details, see \cite{Tegmark1996}.}. 

To simulate such a system we used the so-called cellular automata (CA) models \cite{Ilachinsky2000}. CA models are a common tool to investigate interacting agents in a physical space, which, for our case, will be a surface. Each point in this mesh is a site, which can be either empty or instead occupied by an individual of a given type. 
Let us start with a simple "ecosystem" formed by a species exhibiting two phenotypic traits. Let us indicate as $\Sigma = \{0,1\}$ the two possible "genotypes" which can be understood as two alleles of a given gene. 
\begin{figure*}
\center
\includegraphics[width=1.00\textwidth]{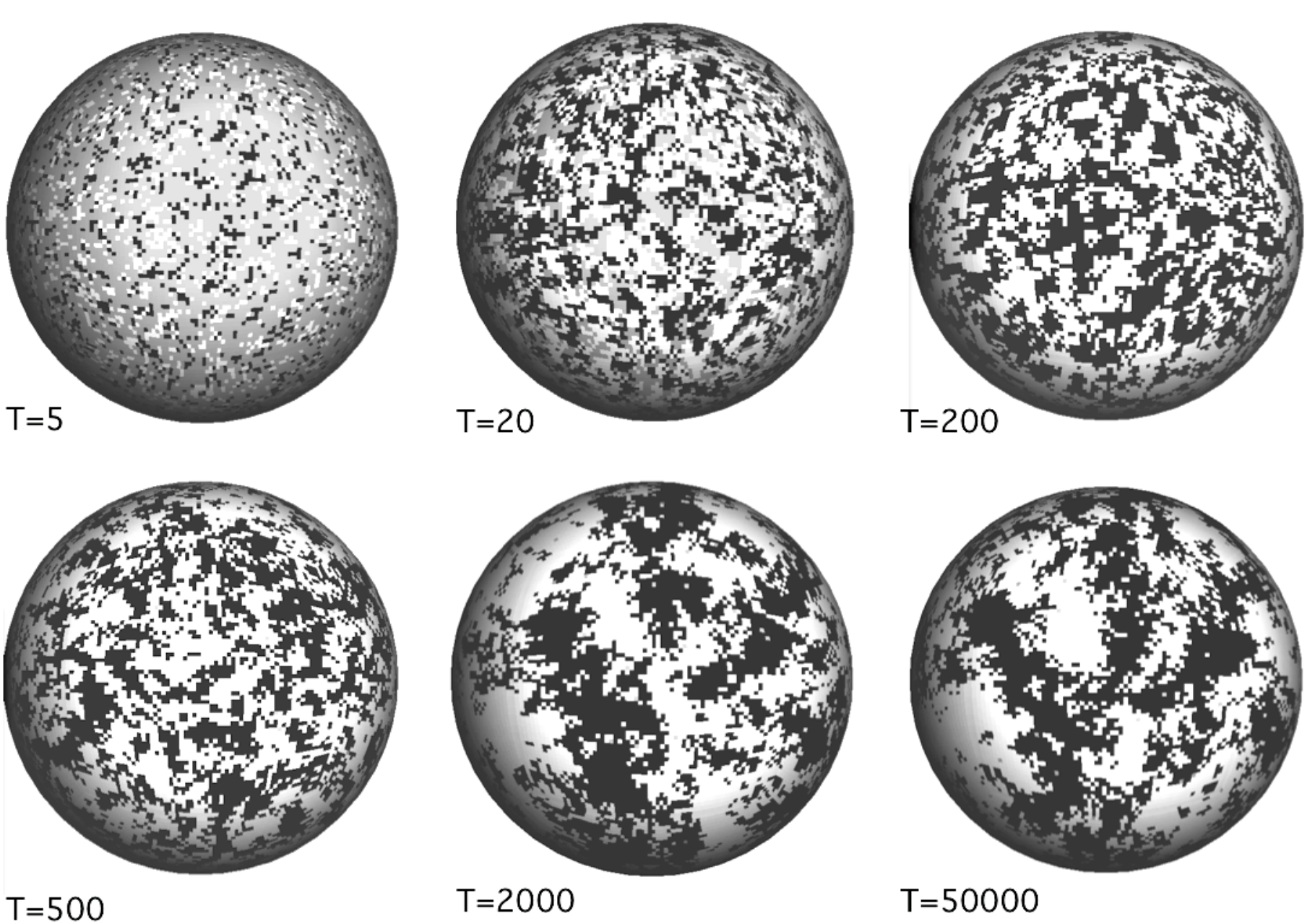}
\caption{Spatial competition dynamics in a single-species model with two mutants. 
Starting from a small set of occupied sites 
with $0$ and $1$ genotypes (white and black, respectively) scattered on an empty (gray) landscape, we display spatial patterns for different simulation times, $T$. 
After a few steps, local populations start to grow, but competition is still weak. Once the two populations occupy 
enough space, competition starts to be effective but no global exclusion occurs. Instead, the two populations coexist 
by expanding locally over spatial domains that appear homogeneous. After a long time, the system is rather 
stable. Although structures keep changing their boundaries, the global picture remains the same, with a spatial 
landscape displaying large homogeneous patches. Here we used $\mu = 10^{-4}$, $\delta_H=10^{-3}$ and $r_H=0.35$.}
\label{figure2}
\end{figure*}

In our idealized model, $0$ and $1$ are the only two genotypes, each one associated to a set of parameters 
defining the underlying phenotype. For simplicity, we will consider completely symmetric sets. In other words, 
we have a neutral change when moving from $0$ to $1$ and viceversa. These transitions occur proportionally to mutation probability $\mu \in [0,1]$.

We can use the previous system to simulate host-parasite interactions with matching alleles (MA, see Section 4 for further details) interactions. Our CA is thus given by a two-dimensional state space, $\Omega (i, j)$, with spatial coordinates $(i,j)$. The states, $S$, of the automaton at time $t$ are given by $S(i,j; t) \in \Sigma =\{H_k, P_k, E\}$, where $H$ and $P$ denote, respectively, hosts and parasites defined as 1-bit strings (i.e., with $k = 0,1$). $E$ indicates empty sites in the state space $\Omega$. 
\begin{figure*}
\center
\includegraphics[width=1.00\textwidth]{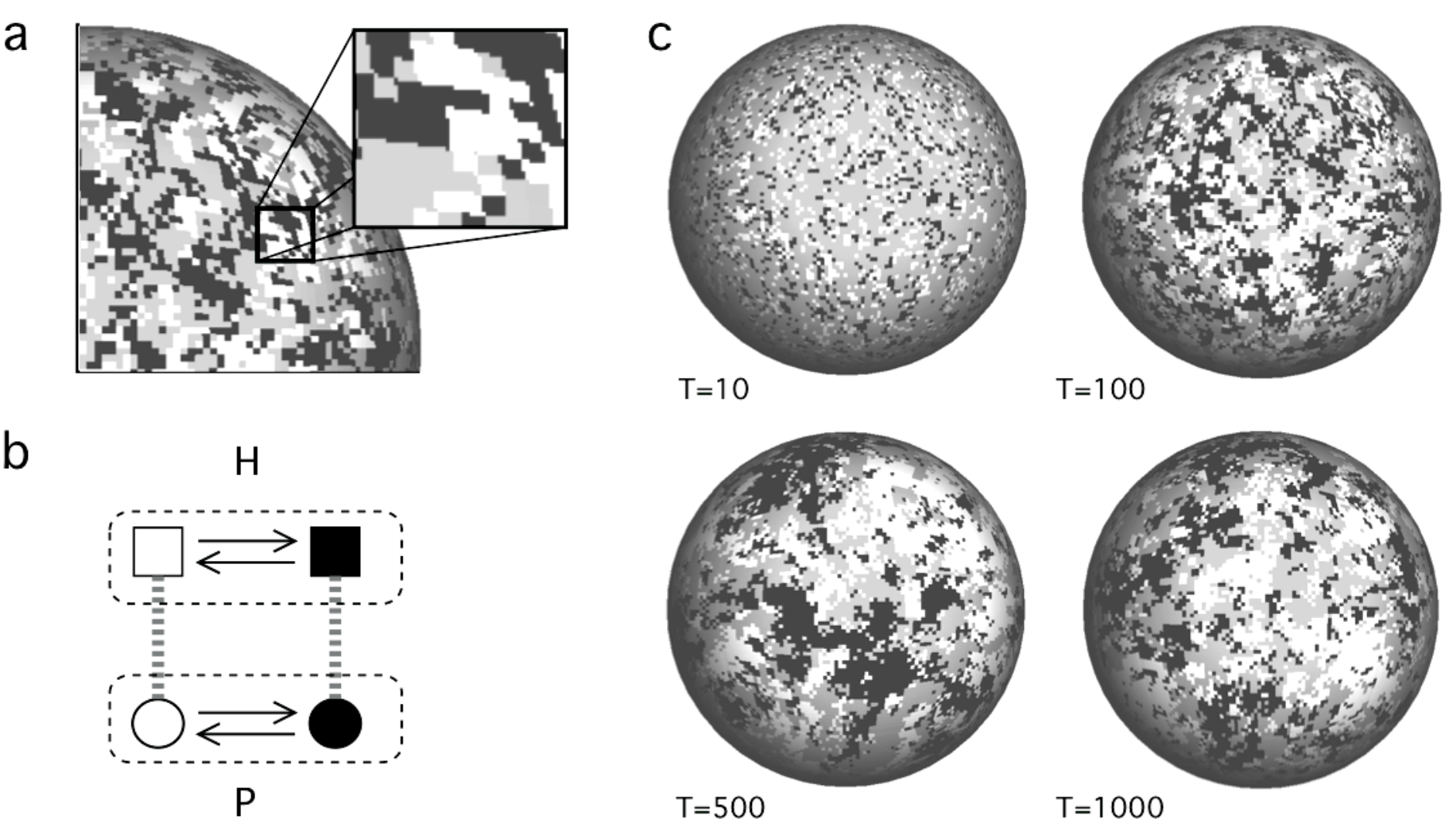}
\caption{Red Queen dynamics in host-parasite (prey-predator) coevolution. Here we have 
added to the only-host system displayed in the previous figure an evolving $1$-bit parasite. 
Parasites propagate through space provided they find the same host genotype. Here 
two levels need to be considered: the spatial framework (a) defined by the local arrangement of 
individuals and the required genotypic matching. (b) Schematic diagram of the 
two populations (upper and lower layers) with back and forth mutations (arrows) between genotypes 
and the requirement of allele matching (vertical lines). The plots in (c) show the 
host populations (as before) which now keep changing. Similar plots would be observed (although having less dense patches) 
for parasites. Here we used, as before, $\mu_H=10^{-4}$, $\delta_H=10^{-3}$, $r_H=0.35$; and $\mu_P=10^{-4}$, $\delta_P=0.2$, and $r_P=0.1$.}
\label{figure3}
\end{figure*}

CA models include dynamics by means of the state-transition rules, which determine the changes of the states according to the current states in a given site together with its neighboring states. For this particular system we used the so-called Moore neighborhood, which considers the eight nearest neighbors. The model rules are summarized in Box 1.
\begin{svgraybox}
{\bf{Box 1.}} State transition rules of the CA model for host-parasite matching allele interactions with $1$-bit genotypes.
\begin{enumerate}
\item
{\bf{Death}}: hosts and parasites decay with probabilities $\delta_H$ and $\delta_P$ respectively, according to (recall $k = 0, 1$):
\begin{equation}
\nonumber{H_k \xrightarrow{\delta_H} E,}
\end{equation}
\begin{equation}
\nonumber{
P_k \xrightarrow{\delta_P} E}.
\end{equation}

\item
{\bf{Birth of new hosts}}: hosts can replicate with probability $r_H$, without and with mutations, following the reactions:
\begin{equation}
\nonumber{
H_k + E \xrightarrow{r_H (1-\mu_H)} 2 H_k.}
\end{equation}
\begin{equation}
\nonumber{
H_k \xrightarrow{r_H \mu_H} H_k + H_{1-k},}
\end{equation}
where $\mu_H$ is hosts mutation rate.

\item
{\bf{Predation}}: Predator genotypes predate on hosts same genotype (assuming a perfect MA interaction), with rate $r_P$. In 
other words, they can reproduce only under the presence of the same host genotype in the neighborhood. If such condition is met, 
the new reactions follow:
\begin{equation}
\nonumber{
P_k + H_k \xrightarrow{ r_P (1-\mu_P)} 2P_k,}
\end{equation}
\begin{equation}
\nonumber{
P_k + H_k \xrightarrow{r_P \mu_P} P_k + P_{1-k},}
\end{equation}
introducing again mutational changes proportionally to parasites mutation rate $\mu_P$. Because of the H-P interaction, hosts experience a parasite-driven 
mortality. 
\end{enumerate}
\end{svgraybox}

What is the dynamics of this simple model when no parasites are present? In figure \ref{figure2} we see an example of 
the time evolution of this model, starting from an initial condition where we scatter a small population 
of each genotype over the sphere. The empty, available space is indicated in gray, whereas the two alternative 
genotypes are shown as black and white squares. After a short transient, where both variants expand 
with no special constrains, available sites become scarce and the expanding patches grow and develop 
rugged boundaries. Such pattern stabilizes in the long run, where we observe large domains of each class. 
This phenomenon is due to the local exclusion of our identical competitors \cite{SoleBascompte-SO} that allows global 
coexistence to occur.

The previous scenario shows that competing populations end up in a predictable community structure with 
no further (global) changes, However, when an additional component -parasites (or predators)- is added in the same system, 
it immediately triggers the emergence of an unstable dynamical state. This is shown in another example in figure \ref{figure3}, 
where we again represent the host populations, now starting with the same condition as in figure \ref{figure2} 
but adding also some randomly distributed predators.

At any step, the spatial distribution changes rapidly and complex waves of 
expansion and contraction, affecting both genotypes, are observable. Sometimes, the extinction of the parasites returns 
our system to the previous conditions without parasite (and a spatial segregation pattern). This occurs for example when 
the mutation rate of the parasite is too small or its death rate too high. Sometimes, the pressure of 
the parasites is so strong that they cause the extinction of the hosts and the eventual collapse of the whole host-parasite  
system. The interesting dynamics occurs when parasites are able to reliably match their preys and reproduce at a reasonable 
pace. Similarly, mutation rates need to be high enough to react to depleted host populations and at the same 
time allow for a conservation of genetic information, thus avoiding undesirable drift (see Section 5.1). In other 
words, intermediate rates of parasite pressure end up in Red Queen evolution.

Here the system constantly changes as a consequence of the hide-and-run effect induced by the parasitic species. 
Each time a parasite finds a suitable host sharing the same bit, it replicates and will keep expanding provided that 
it finds additional hosts in the neighborhood. The end result is a spatial pattern of propagating patches and constantly 
changing distributions of the two available genotypes.

\section{Fitness Landscapes}

Adaptive or fitness landscapes are a very useful tool and an important concept in evolutionary biology. They are used to map, represent or visualize the relationship between genotypes (or phenotypes) and reproductive success or fitness. Fitness landscapes were first introduced by Sewall Wright and were then extended by other 
authors (see \cite{Kauffman1993,Perelson1991} and references therein). Fitness landscapes assume that each genotype has a well-defined fitness value, which is represented with a height or peak in the landscape (see figures \ref{figure5} and \ref{figure8}). A landscape simply means a single-valued scalar function, $F(x)$, of the state or configuration $x$ of a system. The variable $x$ typically has very many dimensions, and thus may be often written like a multidimensional vector, as a set of $N$ components $x_i$: $${\bf{x}} = (x_1, x_2, ..., x_N).$$ The term landscape is inspired from the geographic landscapes in which the height $h$ above sea level is a simple function $h = F(x,y)$, of the two-dimensional location ${\bf{x}} = (x,y)$. In the field of biology, fitness landscapes are generically representing the fitness of a given biological entity as a function of its genotype or phenotype. As biological entity we referee to a given organism or a to particular macromolecule or cell of that individual. 

Fitness is a relative measure, since it may depend on the environment and on other interacting organisms \cite{Perelson1991}. Fitness can be given by several properties, or by a combination of them. For instance, we can use replication or reproduction success as a measure of fitness. Properties like infectivity, migration capacity, ability to cooperate, among others, can also define a fitness which may facilitate survival and adaptation. As Jacob \cite{Jacob1977,Jacob1983} stated, adaptation typically progresses through small changes involving a local search in the space of possibilities (e.g., sequences space). The paradigm is a hill-climbing process via fitter mutants which "move" towards a global or local optimum in the fitness landscape (see figure \ref{figure4}(A)). The hill-climbing framework was originally proposed by Wright \cite{Wright1931,Wright1932}, who introduced the concept of space of possible genotypes. Under this framework, each genotype has a given fitness, being the distribution of fitness values over the space of genotypes a fitness landscape. Depending upon the distribution of fitness values, the fitness landscape will become more or less mountainous. The behavior of an adapting population will depend on how rugged the fitness landscape is, on the size of the population, and on the mutation rate which moves a population from one genotype to another in sequence space. The motion of a population over a fitness landscape also depends on whether the population reproduces asexually or sexually. The latter reproduction involves mixing of genotypes which can involve to reach more distant points in fitness landscape, compared to asexual reproduction \cite{Kauffman1993}. 

The fitness landscape concept has been widely used in both evolutionary and coevolutionary biology. Organisms do not merely evolve, they coevolve with other organisms. As a difference from evolution, that can be roughly characterized as an adaptive search on a "potential surface", in coevolution there may typically be no such potential surface, being the process far more complex \cite{Kauffman1993}. Actually, in coevolutionary processes the adaptive landscape of one organism can deform and heave as the other organisms make their own adaptive moves. Under this perspective, one can interpret coevolution as both dynamical and evolutionary processes occurring in coupled fitness landscapes (see figures \ref{figure3}(b), \ref{figure4}, \ref{figure5}, \ref{figure8} and \ref{figure10}(a)). 

For the sake of clarity, before accounting for coevolution, we will introduce, in Sections 3.1. and 3.2., some information about evolution on fitness landscapes. First, we will describe evolutionary phenomena in simple fitness landscapes also presenting a theoretical body to model evolution in these types of landscapes. Then, we will extend our explanations to more complex, rugged fitness landscapes. From Section 3.3. onwards (together with the Introduction Section above) we will strictly focus on coevolutionary phenomena. As an example of coevolution at small scales, Section 3.4. includes the view of RNA virus evolution from the perspective of the Red Queen hypothesis. The remaining sections will deal with some examples and models in higher biological scales, from complex ecosystems to macroevolution.

\subsection{Simple versus coupled fitness landscapes}

Models on evolution have considered different theoretical and computational frameworks to characterize several levels of complexity. One of the most successful approaches to address evolutionary phenomena (as well as coevolution as we will discuss later) on fitness landscapes is given by the digital genomes approach  \cite{Freund1991}. Under this approach, using as an example the evolution of RNA genomes, we can develop a mapping between RNA sequences (defined as a chain of nucleotides involving 
a four-letter alphabet $\Omega$) and binary sequences, according to:
\begin{equation}
\nonumber{
{\cal F} : \Omega = \{ U, G, A, C \} \longrightarrow \Sigma = \{ 0, 1 \}}.
\end{equation}
Alternatively, one can use another Boolean representation using spins instead of bits:
\begin{equation}
\nonumber{
{\cal F_s} : \Omega = \{ U, G, A, C \} \longrightarrow \Sigma = \{ +1 , -1  \}}.
\end{equation}
Both approaches are equivalent because the mapping has the same nature. However, the spins approach exploits some advantages of considering "up" and "down" configurations to describe the microscopic dynamics.

Let us define the $i$th string of the population, ${\bf S}_i = (S_{i1}, ..., S_{i \nu})$, as digital genomes (i.e., sequences of purines and pyrimidines only incorporating the linear information encoded by the string) of length $\nu$. In order to determine the functional relevance of these sequences, we need to map them to a sequence-fitness measure, which can be defined as:
\begin{equation}
\nonumber{f  :  {\bf S}_i  \in \Sigma^{\nu}  \longrightarrow f ({\bf S}_i).}
\end{equation}
This functional relation can be generically divided into two types: (i) the different bits of a string play an independent role in fitness; or (ii) some bits of the string influence others in a nontrivial way. Case (ii) corresponds to what is known as epistasis. Epistasis occurs when the phenotypic effect of a mutation depends on the presence of other mutations in the genome. Epistasis becomes especially important in highly-compacted genomes that are expected to contain multifunctional proteins or overlapping genes. In this sense, epistasis has been studied and characterized for RNA viruses, both experimentally and theoretically (see \cite{Elena2010} for a review and \cite{Sardanyes2011a,Sardanyes2009}). In a more general way, epistatic interactions play an important role in evolutionary genetic systems almost whenever multi locus genetics matters and plays a central role in the evolution of genetic systems such as sex and recombination, ploidy, genomic segmentation and modularity, genetic incompatibility and speciation, mechanisms of mutational robustness, mutational load for deleterious mutations through genetic drift, and the rate of adaptive evolution \cite{deVisser2007}. 

\begin{figure*}
\center
\includegraphics[width=1.0\textwidth]{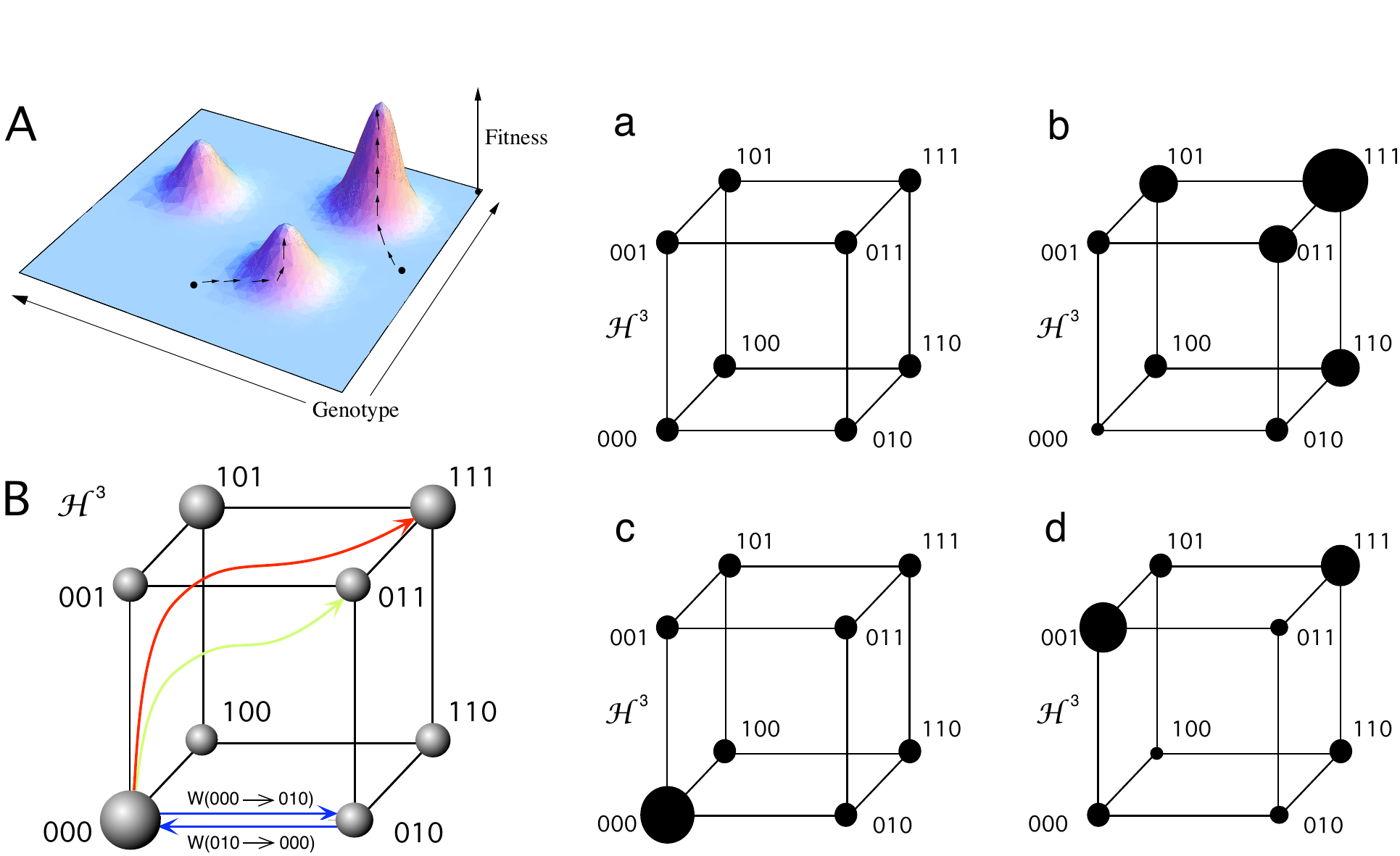}
\caption{(A) Schematic representation of a fitness landscape with three peaks. Depending on the initial condition in the genotype space, the population will evolve towards different maxima due to mutations which result in motions in the landscape. (B) Fitness landscapes can also be represented in the sequences spaces (here using digital genomes). Such systems allow defining trajectories followed through string evolution. Four standard cases of $3$-dimensional sequences space (where the size of the nodes denotes each string fitness value) are also displayed: (a) flat, (b) Fujiyama, (c) Swetina-Schuster (single peak), and (d) rugged fitness landscapes.}
\label{figure4}
\end{figure*}

The digital genomes approach allows us to use an abstract, multidimensional representation 
of the potential set of states accessible to a $\nu$-bits digital genome. This set or space is given by a sequence space in the form of a hypercube, 
$\mathcal{H}^{\nu}=\Sigma^{\nu}$, which can provide, at low dimensions, some intuitions 
about the behavior of strings under selection-mutation pressures (see figure \ref{figure4}). If only small mutation 
rates are considered, transitions between sequences will take place only involving 
nearest neighbors in sequences space, thus differing only in one bit. 

In general, for a given 
mutation rate, $\mu$, two sequences $\bf S$ and $\bf S'$ will be obtained from each other with a given 
probability, given by:
\begin{equation}
\nonumber{
W_{\mu} ({\bf S} \rightarrow  {\bf S}')= \mu^{d_H(\bf S, S')} (1 - \mu)^{\nu - d_H(\bf S, S')},}
\end{equation}

where $d_H(\bf S, S')$ is the Hamming distance among the two sequences (i. e., the number of different bits), with:
\begin{equation}
\nonumber{
d_H(\bf S, S') = \sum_{i=1}^{\nu} \left ( 1 - \delta_{S_i, S'_i} \right )},
\end{equation}
where $\delta_{i,j}$ is Kronecker's delta with $\delta_{i,j} = 1$ if $i = j$ and $\delta_{i,j} = 0$ if $i \neq j$. 
Here $W_{\mu} $ can be interpreted in probabilistic terms: it is the probability of having exactly $d_H$ differences between the two digital genomes.  
This function allows to introduce the dynamics associated to mutations as transition 
probabilities. 
For the spin mapping, the transition probabilities can be expressed as:
\begin{equation}
\nonumber{W_{\mu} ({\bf S} \rightarrow  {\bf S}') = \mathcal{N} \rm{exp} \left(-\beta \sum_{i=1}^\nu S_i S'_i\right)},
\label{spin}
\end{equation}
where the $\beta$ term, defined as $\beta=\log(\mu/(1-\mu))/2$, can be interpreted in terms of 
a temperature, being $\mathcal{N}$ a normalization constant. 

Once we define the fitness function associated to each vertex of the hypercube, we can characterize the dynamics. 

If $N({\bf S},t)$ indicates the fraction of strings having a given sequence ${\bf S} \in \Sigma^{\nu}$ at time $t$, 
Eigen's formulation \cite{Eigen1971} describes the population dynamics as a set of nonlinear differential equations, given by:
%\begin{widetext}
\begin{equation}
{d N({\bf S},t) \over dt} = \sum_{{\bf S}'} W_{\mu} ({\bf S}' \rightarrow  {\bf S}) f({\bf S}')  N({\bf S}',t) -
\left (   \sum_{{\bf S}'}   f({\bf S}')  N({\bf S}',t) \right )  N({\bf S},t) 
\label{eigen}
 \end{equation}
%\end{widetext}

The first term on the right-hand site of Eq. \eqref{eigen} corresponds to positive contributions to the 
abundance of $\bf S$ due to mutation transitions from other strings of the sequence space. The second term 
includes all the reverse events leaving the node occupied by $\bf S$. Figure \ref{figure4}(B) illustrates the information described in 
Eq. \eqref{eigen} for $3$-bits strings (i.e., $\nu=3$). The nodes of the hypercube indicate the population size and the three potential transitions from $000$ to other strings differing in one, two or all bits are indicated by arrows of different colors (see \cite{Elena2010} for the extension of the previous results to discrete dynamical systems). 

Simple fitness landscapes can be defined from our previous definitions. Roughly, a sequences space is a discrete space including all sequence combinations of a given sequence, which are connected to neighboring sequences differing in one bit. This space thus results in a set of nodes or vertices (sequences) which are connected by single-point mutations (see figure \ref{figure4}(a-d)). For a given sequence of length $\nu$, the dimension of the sequences space is given by $\mathcal{H}^{\nu}$. For DNA or RNA sequences, the total number of nodes in the sequences space will be $4^{\nu}$, which results in an astronomic number. As we saw before, this inherent complexity can be reduced using digital genomes defined as bit strings or as "up" and "down" spins. The bit-strings approach allows one to simulate the processes of (co-)evolution and selection under different types of rules or interactions describing different processes in  biological systems. For instance, in cancer dynamics \cite{Sole2003}, in RNA viruses \cite{Sardanyes2011a,Sole1999,Lafforgue2011,Elena2010,Sardanyes2009,Sardanyes2008,Sole2006}, and, in the context of coevolution in matching-alleles dynamics (see Section 4), among others. As a simple, illustrative example, figure \ref{figure4} shows four different simple fitness landscapes. The sequences space in figure \ref{figure4}(a) is a flat fitness landscape, where all the sequences have the same fitness $f_0$, according to $\mathcal{H}(\vec{S}_i) = f_0$. If the sequences space has a least fit sequence (e.g., $0$) and then the fitness increases at increasing number of mutations, we have the so-called Fujiyama fitness landscape, shown in figure \ref{figure4}(b), being its fitness given by $\mathcal{H}(\vec{S}_i) = f_0 + \sum_{k=1}^\nu S_{ik}$ (see \cite{Quer1996} for the application of the Fujiyama landscape to RNA viral populations). Another widely studied case is the so-called single-peak fitness landscape, which is shown in figure \ref{figure4}(c). For this landscape, the fitness can be defines as $\mathcal{H}(\vec{S}_i) = f_0 \delta_{\vec{S_i},\vec{1}}+ f_{1} (1 - \delta_{\vec{S_1},\vec{1}})$, with $f_0 > f_1$ (see \cite{Sole2006,Sardanyes2010,McCaskill2001} for some examples and applications of the single-peak fitness landscape).

\subsection{Evolution on rugged fitness landscapes}

The sharp, single-peak fitness landscape cited at the end of the previous section defines an extreme in a hierarchy of models introducing 
different levels of dependencies among genes. A different approximation deals with 
landscapes in a much more general way, by allowing them to display 
a given number of local maxima generating a mountainous landscape. Together with the simple fitness landscapes we display a rugged fitness landscape in figure \ref{figure4}(d). For this landscape, where each sequence has a different fitness resulting in as many peaks as sequences, the fitness can be given by 
$\mathcal{H}(\vec{S}_i) = \frac{1}{\nu} \sum_{k=1}^\nu \xi_{ik}$, being $\xi_{ik} \in k[0,1]$, a random number. The best known model for the evolution on rugged fitness landscapes is Kauffman's NK model \cite{Kauffman1993,Kauffman1991}, which is also defined on a hypercube. It was originally proposed as a representation of 
haploid genomes involving two alleles per locus with additive contributions 
to fitness from different loci. 

The $NK$ model is a simple model of random epistatic interactions. In this model $N$ is the number of parts of a system (e.g., genes in a genotype or amino acids in a protein). Each part makes a fitness contribution which depends upon that part as well as upon $K$ other parts among the $N$. Thus, $K$ reflects how richly cross-coupled the system is indicating how many other genes influence a given gene i.e., the richness of epistatic interactions among the genes. Such a model, under parameters alteration, generates a family of increasingly rugged multi peaked landscapes. Once again a fitness function is introduced,  ${\bf{f}} = f(S_{i1}, ..., S_{i\nu})$, and changes in the traits are assumed to occur by means of single, one-bit steps (i.e., single-point mutations). This single-chain 
events are consistent with our assumption of small mutation rates. In this way, a given 
string obtained by inaccurate replication allows to perform a random {\em adaptive walk} from a given node towards one of its
$\nu$ nearest neighbors if this leads to an increase in fitness. A direct consequence
of this process is that once a local peak is reached, no further changes are
allowed to occur. This is completely different from the assumptions made above, which assume 
the presence of a preferred sequence around which other sequences have a lower fitness value. 
In the context of NK landscapes, a local peak is very simply defined: if all nearest neighbors in
the hypercube are less-fit, we have a fitness local maximum. 

How can we construct a system displaying a NK landscape? Kauffman suggested a simple 
approach using fitness tables: for each element $S_{ij}$, if it is influenced by $K$ other elements, each 
element contributes in an additive way to the overall fitness. In other words, if we consider the 
two-locus model and assume that a given locus $i$ constributes to the global fitness associated to $\bf S$ 
by an amount $f_i({\bf S}) \in [0,1]$, the global fitness is given by  the average value:
 \begin{equation}
 \nonumber{
f({\bf S}) = {1 \over \nu}  \sum_{i=1}^\nu f_i({\bf S})}.
 \end{equation}
As $K$ grows, the ruggedness of the landscape increases, since more complex interactions 
are allowed to occur.

An interesting feature of the NK model is that, because of its simplicity, it allows 
the prediction of some evolutionary dynamical patterns. As an example, let us 
consider that fitness values are random and uncorrelated, i. e., $f(S_{i1}, ..., S_{i\nu}) = \xi$, where $\xi \in [0,1]$ is a random number with uniform distribution. This random fitness landscape
has many local fitness peaks. This number $M_L$ is very large: 
 \begin{equation}
\nonumber{
M_L(\nu) = {2^{\nu} \over \nu +1}},
 \end{equation}
and thus our digital sequences can get trapped in a very large number of optima. 
To see this, let us consider the number of neighbors of a given
node and compute the probability that this node is a local maximum.
The chance that it is the fittest among its $\nu$ neighbors and itself,
given the random choice of values, is simply $P_1=1/(\nu+1)$. Since there are
$2^{\nu}$ possible strings, the fraction of those who are local maxima 
is $M_L(\nu) =2^{\nu} P_1$. An extension of this model can be easily introduced by means of the so-called Fujiyama landscape (see previous Section), where a 
fitness function is defined now as follows: $$f(S_i) = \frac{1}{z} (1-s)^k,$$ with $k = \nu - \sum_{l=1}^\nu S_{il}$, being $z = \sum_{j=1}^\nu (1-s)^j$ a normalization factor. The parameter $s$ weights the steepness of the peak.

Similarly, the presence of epistatic interactions can be introduced using the sequence-dependent fitness:
$$f(\vec{S}_i) = \frac{1}{2}\left[\left(\frac{1}{2}\right)^{d_H(\vec{S}_i,\vec{S}_n)^\xi} + 1\right],$$
where $\xi >0$ defines the degree and type of epistasis (for $\xi < 1$ we would have antagonistic interactions whereas $\xi > 1$ defines synergistic epistasis (see \cite{Elena2010} and references therein).

\subsection{Coevolution on rugged fitness landscapes}

The previous sections have been developed to provide the reader with a broad framework that will be, from now on, extended to the subject of coevolution. As we already 
focused on evolution on rugged fitness landscapes, let us start with coevolution on these landscapes (in following sections we will analyze coevolutionary dynamical models in simpler 
fitness landscapes). In the context of rugged fitness landscapes, the $NK$ model can be modified to analyze evolution between many interacting species, by means of the so-called NKC model \cite{Kauffman1991}. This model introduces a new parameter, named $C$, which denotes the number of couplings between different species (also represented as strings). Now, the fitness of the NK model needs to be modified in order to introduce the coupling effects where each trait receives inputs from other $C$ other traits from different species. These traits are randomly chosen between the $S$ species of the ecosystem. The $NKC$ model includes three main parameters describing: (a) the number of traits required to characterize a given species ($N$), (b) the number of so-called epistatic interactions among genes in the same species ($K$), and (c) the number of interactions among traits of different species ($C$), which introduce coevolution. 
\begin{figure*}
\center
\includegraphics[width=.67\textwidth]{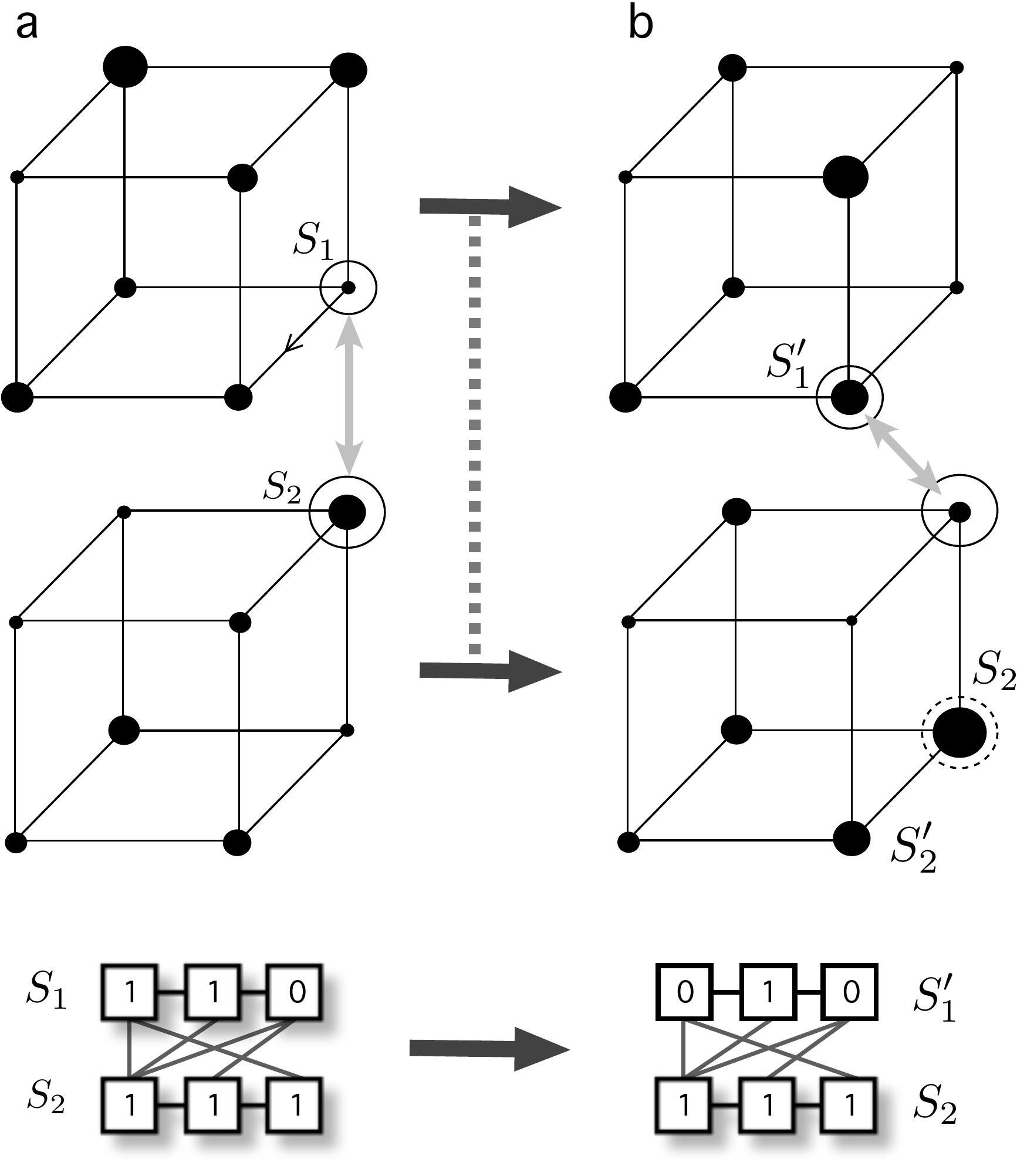}
\caption{Coevolutionary dynamics in the NKC model. Here only two species ($S_1$ and $S_2$) are considered, each one described by $N=3$ traits, which are represented in cubic sequences spaces. Black circles in the hypercube nodes indicate the fitness associated to 
each string, and current states are highlighted with open circles in (a). Each bit is assumed to interact with 
a number of different bits from the other string (genome) as indicated in the lower diagrams. Once the first species changes by 
climbing to a local optimum, the landscapes of both species get modified. The second species now will be forced 
to change too, since it is now placed in a low-fit state and will next shift to another local peak (here 
indicated with a dashed circle in (b)). If no such movement is possible, extinction can take place. }
\label{figure5}
\end{figure*}

Figure \ref{figure5} illustrates this approach for $N = 3$ traits of two interacting species. In the figure, the local peaks are indicated by black circles. Each species is defined by a set of traits, which are coded by bit-strings. Such traits are connected among the different species. In figure \ref{figure5}(a), species $1$ is not located in a local peak. As a result of an adaptive walk, it will reach the local peak by mutation. However, as a result of the change in species $1$, species $2$ is now not located in a fitness peak. The landscape of species $2$ has been modified by the adaptive motion of species $1$.

The $NKC$ model was analyzed computationally by Kauffman and Johnsen \cite{Kauffman1991}, and they showed that this system was dynamically very rich. They identified a chaotic phase, where the ecosystem is always changing and species never end up in a particular configuration (i.e., species do not stop at a given local peak). As we will discuss later, many other different models suggest that chaos can be found in Red Queen dynamics. They also identified a frozen phase, where all species settle down to local peaks. Interestingly, for finite systems at the boundary between the chaotic and the frozen phase in the parameters space, small perturbations generate a coevolutionary avalanche of changes through the system. Typically occurring at critical states, the distribution of such avalanches was shown to follow a power-law. Kauffman and Johnsen mapped these avalanches into extinction events, suggesting that the number of changes in species was proportional to the extinction of less-fit variants. Such a result did not fit the predictions of fossil record extinctions. However, a variation of this model by Newman and Palmer \cite{Newman2003}, which allowed changes in the parameters, gave an exponent which agreed with fossil record data (see also Section 6).

The two phases of the $NKC$ model can be derived from simple theoretical arguments setting $K = N-1$ \cite{Bak1992}. It is known that a given species, in order to reach a local fitness, needs a number of walks $L_w$, which is on average $L_w \approx \ln(N)$. If we assume that all species are at local peaks and one of them, named $a$, is perturbed (i.e., is randomly positioned on the fitness landscape), then $a$ will start again to climb some other local peak. If $C$ is large enough (i.e., interactions among species are important), the other species except $a$ can see their landscape modified also starting to change. Following this idea, where each adaptive walk involves a change in a given trait, which can in turn affect other species traits, we can determine a critical condition given by a combination of $N$ and $C$ able to trigger a chain reaction able to percolate through the system. More concretely, the probability that a given trait in a random species depends on species $a$ is $C/N$. The critical condition is that at least one change in a species occurs. This actually means:
\begin{equation}
\nonumber{L_w \frac{C_c}{N} = 1,}
\end{equation}
being $C_c = N/ln(N)$. That is, when, on average, one out of $C$ randomly chased genes is among the $L_w$ changed genes. In other words, when the number of traits is such that $C > C_c$, interactions among different genotypes constantly modify the underlying fitness landscapes, scenario under which coevolutionary avalanches take place.

\subsection{Red queen dynamics in RNA virus}

Let us first consider a very simple example of coupled fitness landscapes and two identical populations climbing and competing on them. The Red Queen hypothesis of evolution has been widely discussed within the context of RNA viruses \cite{Quer1996,Sole1999}, where the dynamics of viral populations can be interpreted as a dynamical process of growth, competition and selection taking place in the sequence space. The fitness landscape for a virus is usually defined in terms of replication rate or infectivity or transmission. The landscape appears as a multipeaked surface, where the local maxima represent optimal fitness values which can be reached by mutation. Here, the initial condition plays an important role since depending on where the quasispecies\footnote{The term \emph{quasispecies} is used to define the heterogeneous population of viral genomes in RNA viruses.} is located in the sequences space, the population will evolve by exploring near genotypes by mutation. Competition experiments between several clonal viral populations \cite{Clarke1994,Quer1996} provided a good illustration of two basic principles of evolutionary ecology: the Red Queen dynamics and the principle of competitive exclusion.

\begin{figure*}
\center
\includegraphics[width=1.00\textwidth]{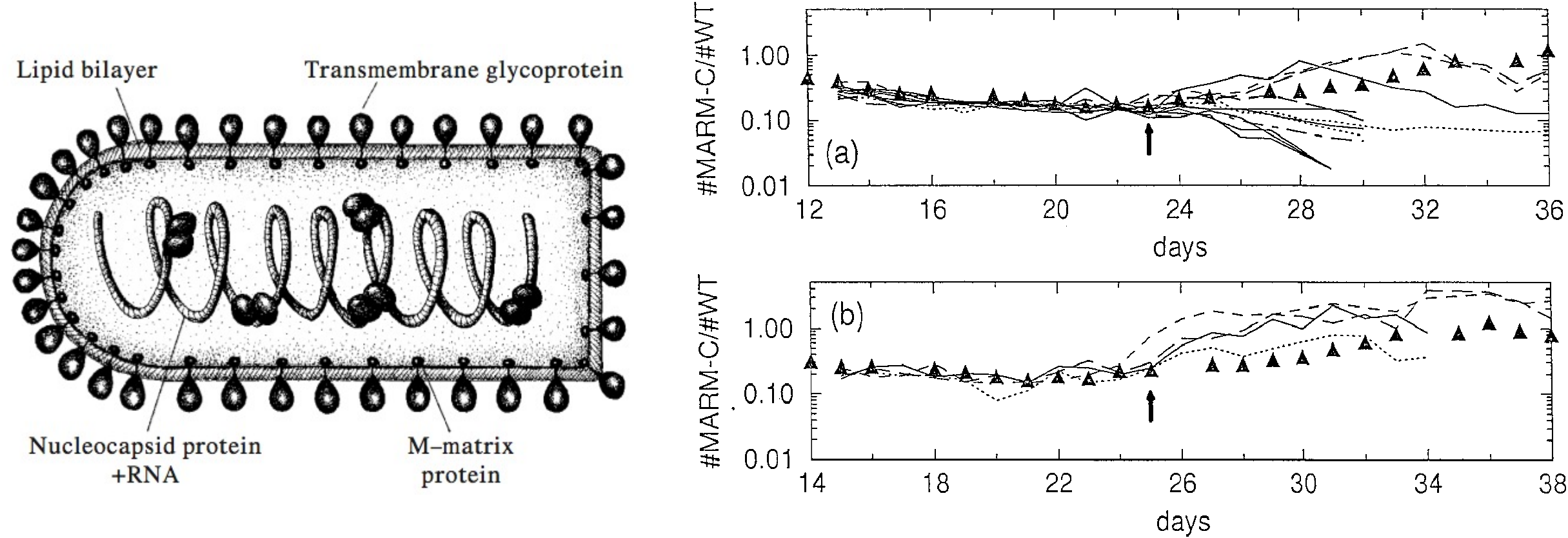}
\caption{Left: Vesicular stomatitis virus (VSV) virion which contains a negative-sense, single-stranded RNA genome. 
The bullet shape is characteristic of the \emph{Rhabdoviridae} family (drawing by Ricard V. Sol\'e). 
Right: Time changes of MARM-C:wt ratio in independent replicas. (a) Shows eighteen replicas started after passage $12$. (b) Displays six replicas started from passage $14$. In both panels the populations start to diverge after approximately $23$ passages (indicated with small arrows). Plots obtained from \cite{Quer1996}.}
\label{figure6}
\end{figure*}

Experimental results were carried out with two clones of Vesicular stomatitis virus (VSV, see figure \ref{figure6}). Such experiments involved the mixing  of two clonal populations of VSV of equal fitness. Passage experiments allowing these populations to grow and compete were performed using standard virus plaque assays. 
More specifically, genetically marked monoclonal antibody-resistant (MARM) clones of equal fitness to the wild-type VSV were used and their relative frequencies were monitored along passages. 

The MARM clones only differed in a single mutation with neutral affects not changing viral fitness. These experiments revealed that both competing populations grew up showing steady increases of fitness, but, at some point, one of the two populations suddenly excluded the other one. The winner of this competition process was not always the same (see figure \ref{figure6}). Although the time scale of the divergence seemed highly predictable. The simultaneous increase of fitness of the two populations and their predictable divergence was suggested to be a product of the Red Queen effect \cite{Sole1999}. In the context of RNA viruses, newly arising mutants with higher fitness were able to outcompete lower-fitness ones. At the level of viral genomes or sequences, a favorable mutation within one quasispecies triggers evolutionary responses in the second one, forcing it to evolve. Overlapped with this evolutionary process, and related to the dynamics, the principle of competitive exclusion is also at play. This principle states that when two species are strongly competing for the same finite resources, the fitter one asymptotically outcompetes the least fit. 
\begin{figure*}
\center
\includegraphics[width=0.79\textwidth]{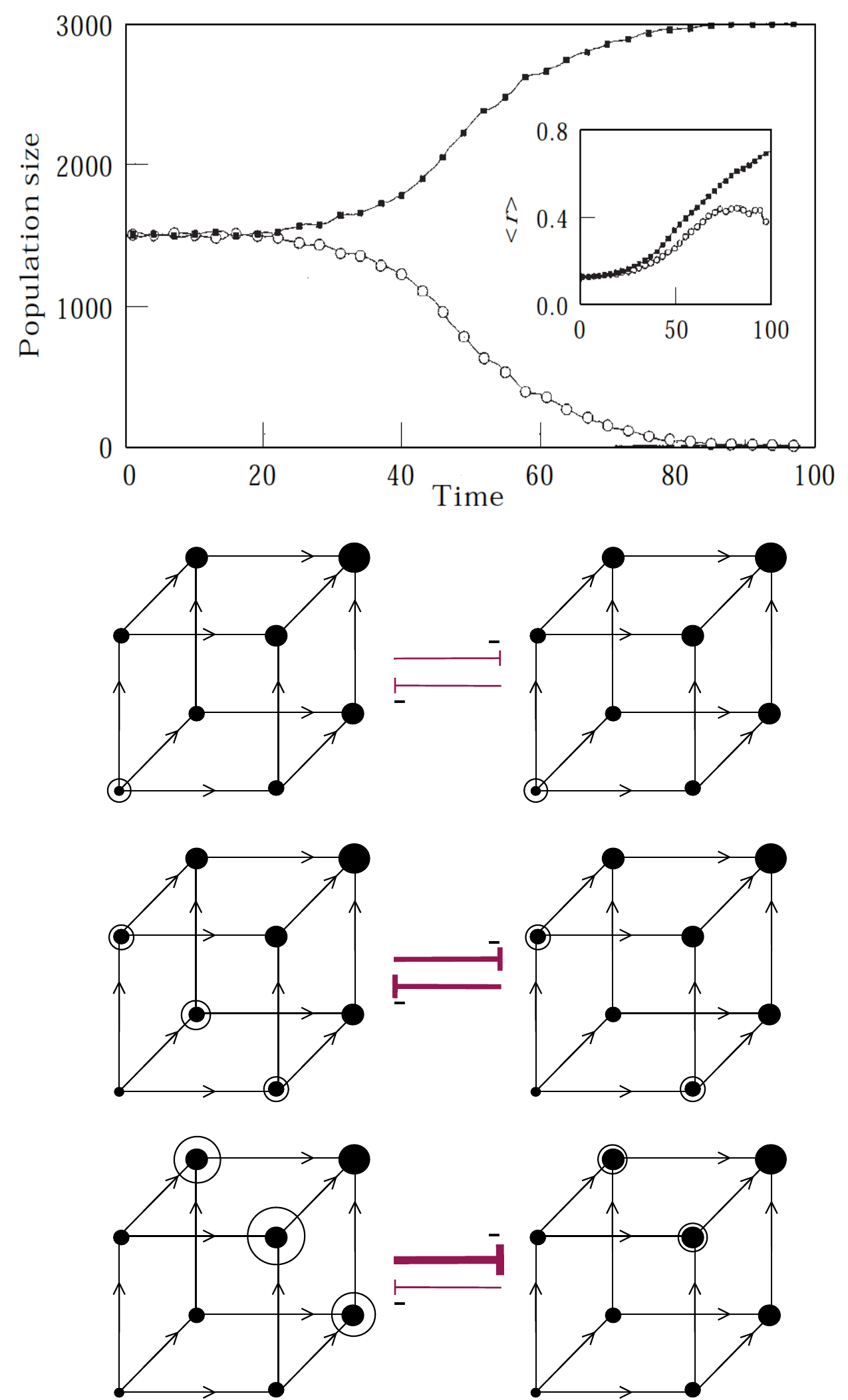}
\caption{(Upper) Dynamics of the bit strings model using a population with $N = 3000$ strings of length $\nu = 16$ using $\mu = 10^{-3}$. The main plot shows the time dynamics of both populations. The inset shows the mean fitness, $\left < r \right >$, also for both populations along time (see \cite{Sole1999}). The observed changes can be easily interpreted in terms of a parallel climbing of both species on their 
Fujiyama landscapes together with ongoing competition for resources. Below we illustrate this by means of a small, three-bit landscape. Initially 
both species (their populations are indicated with open circles) grow slowly and competition is weak. As they climb up and 
increase their replication rates, competition become strong and symmetry breaking occurs (see text). }
\label{figure7}
\end{figure*}

The previous experiments were modeled by Sol\'e and collaborators \cite{Sole1999} using different approaches. The simplest one was a bit-string model that considered a population of $N$ bit-strings, named $S_i$, with sequences: $S_i = S_i^1 S_i^2...S_i^{\nu}$; with $i = 1, 2, ... N$ and $S_i^j \in \{0,1\}$. At each time generation (passage), the algorithm repeated $N$ times the following rules: a random string, say $S_i$, of the population was chosen for replication. Replication, proportional to replication probability $r(S_i)$, took place by replacing one of the strings of the population (also randomly chosen), say $S_j$ by a copy of $S_i$. Replication presented error, at a rate (per bit and replication cycle) $\mu$. So the probability to copy exactly the same bit was $1-\mu$. The mapping between sequence composition (genotype) and replication rate (phenotype) was done using the Fujiyama fitness landscape (see figure \ref{figure4}(b)), involving the linear relation: $$r(s_i) = \frac{1}{\nu} \sum_{j=1}^{\nu}S_i^j.$$ 
As we previously explained, this fitness landscape ignores epistatic interactions. 

Simulations revealed the same behavior obtained with the experiments with VSV. Figure \ref{figure7} displays the outcome of the model for a population of $N=3000$ strings of length $\nu = 16$. The populations were initialized in such a way that the initial fitness of both populations was low, also keeping equal their mean replication rates. During the simulations, the strings were competing for the available space (i.e., $N$ available sites). The upper panel of figure \ref{figure7} shows the total population size of each population, which was maintained roughly constant along time. However, after approximately $t \approx 20$ passages, one of the two populations started outcompeting the other one, that finally disappeared. Once ecological competition became tight, selection pressure became stronger and the initial parallel growth in fitness for both populations was no longer sustainable. This dynamical divergence was a direct outcome of a "symmetry breaking" phenomenon which explained the VSV experiments (see \cite{Quer1996,Sole1999} for details).

\section{Gene-for-gene and Matching Alleles Models of Coevolution}

\begin{figure*}
\center
\includegraphics[width=1.0\textwidth]{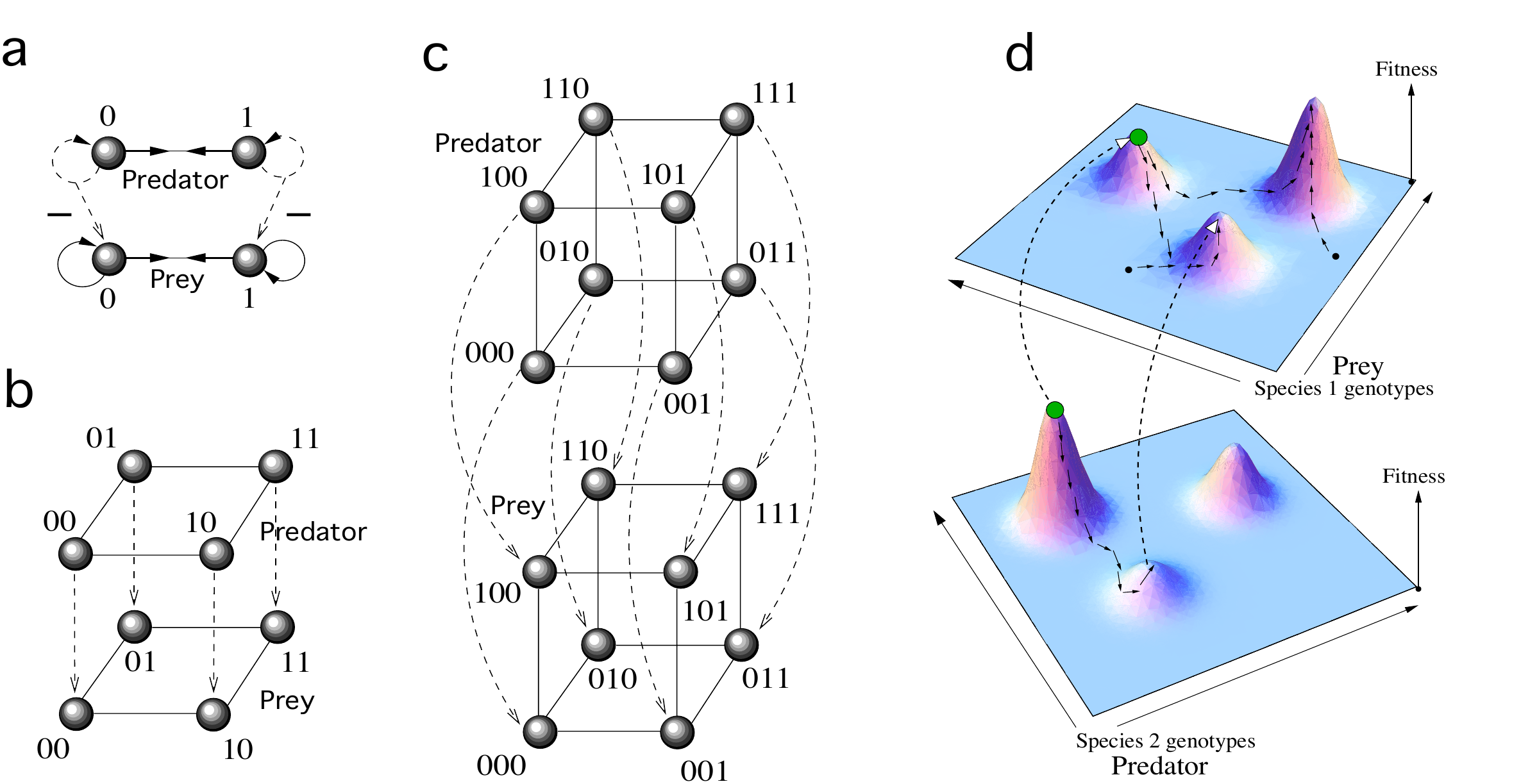}
\caption{Coevolutionary phenomena can be interpreted as the coupling of fitness landscapes by means of ecological interactions (dashed arrows). We show neutral sequences spaces for predator-prey (host-parasite) species with perfect matching alleles interactions for (a) $\nu=1$, (b) $\nu=2$ and (c) $\nu=3$, being $\nu$ the length of the sequences. (d) Prey's evolutionary fate will not only depend on its own and independent exploration of the fitness landscape, but also on predator's evolution (indicated by the small arrows). If a given prey (green circle) moves towards the highest peak, it will escape predator's action increasing its fitness. However, if prey climbs to the lowest peak, and the predator mutates moving to the same peak, host's fitness and reproductive success will diminish.}
\label{figure8}
\end{figure*}

Coevolution is the change of a biological object triggered by the change of a related object. Coevolution can occur at many biological scales: at the molecular level as correlated mutations between amino acids in a protein \cite{Yip2008}, or at the macroscopic scale as covarying traits between different interacting species in an environment. In coevolution, each entity exerts selective pressures on the other, thereby affecting each other's evolution. This process is schematically illustrated in figure \ref{figure8}(d). Here we show two fitness landscapes for preys and predators. Imagine preys are viruses (or cells infected by viruses) and predators are cytotoxic lymphocytes (cells of the immune system that kill infected cells) that act upon the activation of the adaptive immune response. If the virus, located in the peak with the green dot is able to mutate, visiting the highest peak, the immune system will not be able to recognize and remove it. However, if the virus moves towards the lower peak, which is recognized by dendritic cells or macrophages, able to trigger the immune response, virus populations with this genotype will be impaired in terms of number of particles due to the action of cytotoxic lymphocytes, that will remove infected cells. This simple example illustrates how the evolution of one of the partners influences the evolution of the other, and viceversa, in a coevolutionary arms race. Broadly speaking, coevolutionary interactions can be antagonistic or mutualistic. The former involve negative interactions such as predation or parasitism. The latter being found when two or more species coevolve by means of cooperation. Coevolution can occur for two interacting species (pairwise coevolution) or can involve a number of different species, which are evolving in responses to another set of species (diffuse coevolution). 

There are at least about six proposed forms of coevolution between species, some involving reciprocal adaptation and others a combination of adaptation and speciation \cite{Thompson1982,Thompson1989,Thompson1990}. In the context of coevolution between hosts and parasites, some of them had a particular importance in this type of interactions. One is Ehrlich and Ravens \cite{Ehrlich1964} hypothesis of how the evolution of defence and counterdefence in host-parasite interactions may lead to the radiation of species through the process of escape-and-radiation coevolution.  The genetic basis of infection in real ecosystems has been also represented by two major models: the so-called gene-for-gene (hereafter GFG) and matching alleles (hereafter MA) models. The GFG model is based on data from plant-pathogen interactions, especially in the field of crop plants \cite{Flor1956}. Interestingly, the first mathematical model of coevolution was explicitly based on assumptions of a GFG interaction \cite{Mode1958}. Later, a multitude of real examples on GFG coevolution were identified between plants and pathogens, mainly between plants and fungi, bacteria and viruses (see \cite{Thompson1992} for a review). The GFG hypothesis states that "for each gene that conditions reaction in the host there is a corresponding gene that conditions pathogenicity in the parasite" \cite{Thompson1992,Kerr1987}. The key feature of this model is that one parasite genotype can infect all host genotypes.

As a difference, in the MA model, favored by invertebrate zoologists \cite{Grosberg2000}, an exact genetic match is required for infection (figure \ref{figure8}(a-c)). MA models underlie most of the theory constructed to understand the effects of host-parasite coevolution on sex and recombination \cite{Hamilton1990,Howard1994}. Parker \cite{Parker1994} pointed out the importance of MA models for the study of sexual reproduction, suggesting it may hamper the generality of the Red Queen theory for sex. It has been argued that both GFG and MA models are not totally disconnected: they are two ends of a continuum (see \cite{Agrawal2002} for further details).

In the following sections, we will introduce and review recent models and results about coevolving replicators with antagonistic interactions, mainly focusing in MA models. These models, although being suitable to analyze coevolution at small scales (e.g., immune system-viruses) provide good intuitions in larger scales, such as in spatially-extended ecosystems (Section 5.1.). In section 5.2. we will develop some theory aimed to describe the dynamics arising from MA predator-prey interactions. Finally, we will explore large scale coevolution by means of a complex network model reproducing the extinction pattern found in the fossil record data discussed in the Introduction Section.

\section{Minimal Coevolutionary Systems}

Coevolutionary phenomena can be studied considering minimal models. Such models can help us to understand fundamental phenomena arising from species interactions and evolution. We notice that coevolution is a highly nonlinear phenomena, since species interactions give place to nonlinear couplings that can result in very rich and complex dynamics. In this section we will first introduce a minimal system of replicators with matching alleles (MA) dynamics moving, replicating and evolving on a surface. Then, we will develop a general mathematical model describing  MA interactions for haploid genotypes, assuming well-mixed populations thus ignoring spatial correlations. As the reader will see, the dynamics of such small and simple systems can indeed be very complex.

\subsection{Spatial Red Queen dynamics}

At the beginning of this chapter we have illustrated the idea of coevolution with a very simple model simulating replicator spatial dynamics with MA interactions. Together with such model, other approaches have focused on the same subject by considering further complexity, such as spatial diffusion of replicators or larger sequences spaces. Recently, Sardany\'es and Sol\'e \cite{Sardanyes2007a} explored a similar system simulating coevolution for host-parasite (prey-predator) replicators also using cellular automata (CA) models. 

The authors explored the spatio-temporal dynamics for three different host-parasite systems considering $1$-bit, $2$-bits and $3$-bits strings [the corresponding coupled hypercubes are displayed in figure \ref{figure8}(a-c)]. Thus, one of the aims of their work was to analyze the effects of increasing the size of the sequences space in the spatio-temporal dynamics for MA interactions. 
\begin{figure*}
\center
\includegraphics[width=1.00\textwidth]{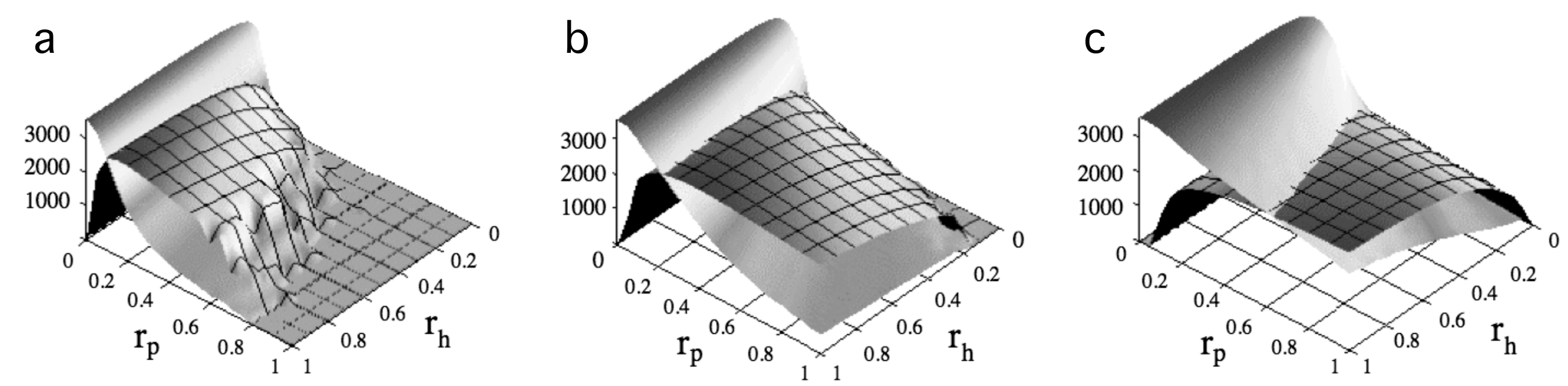}
\caption{Population equilibria for hosts (plain surfaces) and parasites (gridded surfaces) replicators in the parameters space $(r_h, r_p)$ for the spatially-extended model of coevolution with diffusion using $\mu_h = \mu_p = 10^{-2}$. Three different systems were analyzed, with: $\nu = 1$ (a), $\nu = 2$ (b) and $\nu = 3$ (c) (see \cite{Sardanyes2007a} for further details).}
\label{figure9}
\end{figure*}

The model considered two populations of replicators given by host (prey) strings of size $\nu$: ${\bf S}^i_{h} = (s^{i1}_{h}, ..., s^{i\nu}_{h})$; and by parasite (predator) strings of the same size, ${\bf S}^i_{p} = (s^{i1}_{p}, ..., s^{i\nu}_{p})$ with $S^{ij}_h, S^{ij}_p \in \lbrace0,1\rbrace$ where $i = 1, ..., N$, being $N$ the number of different genotypes ($N = 2^\nu$). Both populations reproduced and evolved on a two-dimensional space with toroidal boundary conditions. 
For this model we used the so-called von Neumann neighborhood, which considers interactions with the four nearest neighbors. Specifically, the state-transition rules of the CA considered self-replication with mutation, decay and spatial motion of strings. Rules implemented are shown in Box 2.
\begin{svgraybox}
{\bf{Box 2.}} State transition rules of the host-parasite CA model with matching allele interactions \cite{Sardanyes2007a}: 
\begin{enumerate}
\item
{\bf{Self-replication}}: If a host and a parasite occupy the same spatial position and have the same sequence of bits (i.e., perfect MA), the parasite eliminates the host and replicates, with probability $r_p$, to a random neighbor provided it is empty. If only the host lives in the cell, it replicates with probability $r_h$ to a neighbor cell provided it is not occupied by another host string. Replication involves point mutations for host and parasites, with mutation probabilities $\mu_h$ and $\mu_p$, respectively. These rules can be represented by the following set of reactions:
\begin{equation}
 S^i_h + \vartheta\xrightarrow{r_h(1-\mu_h)^\nu} 2S^i_h,
 \label{efhr}
\end{equation}
\begin{equation}
S^i_h + \vartheta\xrightarrow{r_hW_{ij}^h} S^i_h + S^{j \neq i}_h.
\label{ehr}
\end{equation}
Reactions \eqref{efhr} and \eqref{ehr} denote, respectively, error-free and erroneous host replication. 
\begin{equation}
S^i_h + S^j_p + \vartheta\xrightarrow{\delta_{ij}r_p(1-\mu_p)^\nu} 2S^j_p,
\label{efpr}
\end{equation}
\begin{equation}
S^i_h + S^j_p + \vartheta\xrightarrow{\delta_{ij}r_pW_{jl}^p} S^j_p + S^{l \neq j}_p.
\label{epr}
\end{equation}
Similarly, reactions \eqref{efpr} and \eqref{epr} represent, respectively, error-free and erroneous parasites replication, which is nonlinear due to the density-dependence of the antagonistic interaction.
The parameter $\delta_{ij}$ is again the Kronecker $\delta$ function where $\delta_{ij} = 1$ if $i = j$ and 0 otherwise; and $\vartheta$ indicate some available 
building blocks (i.e. mononucleotides) 
needed to built new strings. The terms 
$W_{ij}^k$, with $k = \{h, p\}$, correspond to the probabilities of erroneous 
replication for hosts ($h$) and parasites ($p$), and are given by:
\begin{equation}
\nonumber{
 W_{ij}^k = (1-\mu_k)^{\nu-d_H[S^k_i,S^k_j]}\cdot\mu_k^{d_{H}[S^k_i,S^k_j]}},
\end{equation}
being $d_H[S^k_i,S^k_j]$ the Hamming distance between two sequences:
\begin{equation}
 d_H [S_i^k,S_i^k] = \displaystyle\sum_{i= 1 }^\nu \arrowvert s_i^k - s_i^k \arrowvert.
 \label{hd}
\end{equation}
Equation \eqref{hd} is a function returning the number of different bits 
when comparing two sequences. Such a function is also used to determine the matching allele interaction between both host ($S_i^h$) and parasite ($S_i^p$) sequences, now with $d_H [S_i^h,S_i^p]  = \sum_{i= 1 }^\nu \arrowvert s_i^h - s_i^p \arrowvert$, where $s_i^h$ and $s_i^p$ represent the bit value ($0$ or $1$) in the $i$th position in both strings. A perfect matching allele interaction will occur when $d_H [S_i^h,S_i^p]  = 0$.\\

\item
{\bf{Molecular decay}}: Host and parasite strings decay with probability $\delta_h$ and $\delta_p$, respectively, according to:
\begin{equation}
\nonumber{
 S^i_h \buildrel \delta_h \over \longrightarrow \vartheta},
\end{equation}
\begin{equation}
\nonumber{
 S^i_p \buildrel \delta_p \over \longrightarrow \vartheta},
\end{equation}
\item
{\bf{Local diffusion}}:
Host and parasite strings move, independently and randomly, to empty neighbor cells with diffusion probabilities $D_h$ and $D_p$, respectively.
\end{enumerate}

To simplify the analysis, all the simulations were run with maximum diffusion constants $D_h = D_p = 1$, also setting $\delta_h = \delta_p = 10^{-2}$. The lattice was randomly inoculated by either host and parasites random sequences.
\end{svgraybox}

As we previously mentioned, the rules were implemented for three different systems with different strings' lengths: $\nu = 1$, $\nu = 2$, $\nu = 3$. For all three different values of $\nu$, the system underwent the same three types of asymptotic dynamics: (i) stable coexistence of hosts and parasites with sustained fluctuations; (ii) hosts survival and parasites extinction; and (iii) both hosts and parasites extinction (i.e., coextinction). The visualization of the populations trajectories in phase space revealed the presence of chaotic coevolutionary dynamics \cite{Sardanyes2007a} (see also \cite{Sardanyes2011b}). In order to characterize the importance of these three possible asymptotic states listed above, we built parameter spaces considering two key evolutionary parameters of hosts and parasites coevolution: self-replication and mutation rates. Figure \ref{figure9} illustrates the outcome of some simulations in the parameters space $(r_h,r_p)$ showing the average population numbers for both global populations\footnote{by global populations we mean the sum of all possible genotypes for a given population i.e., hosts or parasites.} after some transient was removed for $\nu = 1$ (figure \ref{figure9}(a)); $\nu = 2$ (figure \ref{figure9}(b)); and $\nu = 3$ (figure \ref{figure9}(c)). A key result was that the increase of the length of the replicators (i.e., increase of the size of the sequences space) promoted stable coevolution, as shown by the reduction of host-parasite coextinctions (figure \ref{figure9}). 

Moreover, for each of the genotype lengths we simulated three different scenarios, characterized by different values of hosts and parasites mutation rates. The simulations revealed that asymmetries in mutation rates between hosts and parasites had an important effect in the population dynamics: hosts were only able to escape from parasites (causing parasites extinction) if they mutated much faster. Under this condition, the scenario of host's survival and parasites extinction was found for extremely low values of hosts' self-replication rates. On the contrary, when $\mu_p > \mu_h$, the region of parameters space with host and parasite extinction increased for the three hypercubes analyzed, indicating that when parasites evolved faster than hosts they were more efficient in catching hosts thus increasing coextinction phenomena.

\subsection{Dynamics of small replicators with matching-allele interactions}

\begin{figure*}
\center
\includegraphics[width=1.0\textwidth]{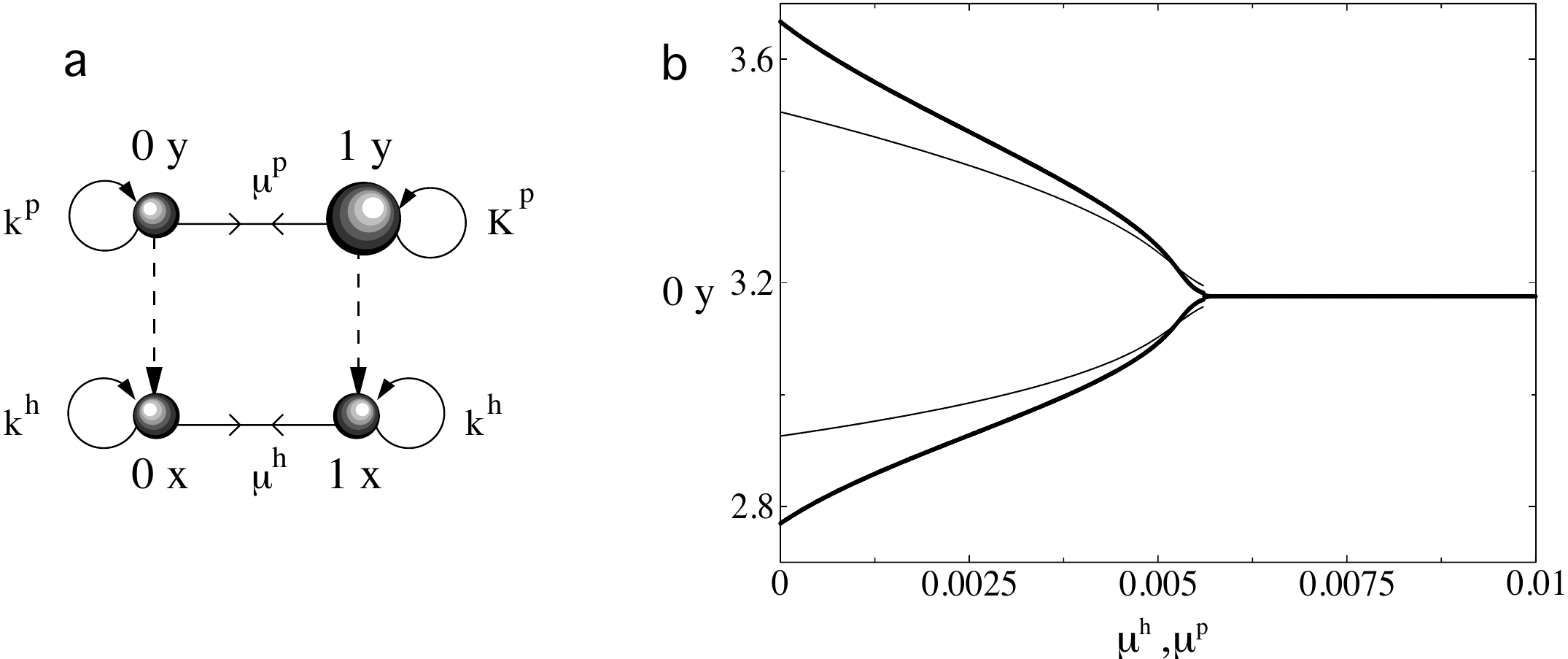}
\caption{(a) Minimal predator-prey system with matching alleles interaction (vertical dashed lines) modeled with Eqs. \eqref{g_RQD1}-\eqref{g_RQD2} for $\nu = 1$. Upper and lower marbles correspond, respectively, to predator ($y$) and prey ($x$) genotypes, which reproduce at rates $k_p$ and $k_h$ (circular arrows), and mutate at rates $\mu_p$ and $\mu_h$ (straight arrows), respectively. We show an asymmetric fitness landscape for predators, with larger replication rates for predator genotype $1$ (i.e., $K_p > k_p$). (b) Bifurcation diagram for predators with genotype $0$, using mutation rates ($\mu^h$ and $\mu^p$ are represented with thick and thin lines, respectively) as control parameters for the neutral fitness landscape (i.e., $K_p = k_p$).. The system undergoes a Hopf bifurcation resulting in a permanent oscillatory behavior governed by a periodic orbit \cite{Sardanyes2008a}.}
\label{figure10}
\end{figure*}

The previous computational models considered antagonistic populations of bit-strings replicating and mutating on a surface. The same system can be investigated with a mathematical model by assuming no spatial correlations (i.e., infinite diffusion). A general model describing predator-prey matching-alleles (MA) interactions can be formulated using a time-continuous deterministic model. Assuming a perfect MA [see figure \ref{figure8} and figure \ref{figure10}(a)], where each predator genotype can predate only on its homologous prey genotype 
(i.e., predator genotype $i$ predates on prey genotype $i$, with $i \in \{0,1\}$), the model is given by the following system:
\begin{eqnarray}
\nonumber{\dot{x_i}=k^h_i x_i \left(1 - \frac{\sum_{j \in \cal H^{\nu}} x_j}{\mathcal{K}}\right) - A_i \xi(x_i,y_i) +} \\
+ \frac{\mu^h_i}{\nu} \left(\sum_{<j>_i} x_j - x_i\right) - \epsilon^h_i x_i,
\label{g_RQD1}
\end{eqnarray}
\begin{equation}
\dot{y_i} = \xi(x_i,y_i) + \frac{\mu^p_i}{\nu}\left(\sum_{<j>_i} y_j - y_i\right) - \epsilon^p_i y_i,
\label{g_RQD2}
\end{equation}
with:
\begin{equation}
\xi(x_i,y_i) = \frac{k^p_i y_i x_i}{C_i + x_i}.
\label{ht}
\end{equation}

The state variables $x_i$ and $y_i$ indicate, respectively, the concentration or 
population numbers of the $i$th prey and of the $i$th predator genotype 
(with $i=1...2^{\nu}$), which define a $\nu$-dimensional sequences space 
$\cal H^{\nu}$. 

Note that prey genotypes have 
a logistic-like growth constraint in self-replication indicated 
in the first term in parenthesis of Eq. \eqref{g_RQD1}, where 
$\sum_{j \in \cal H^{\nu}} x_j$, 
is the total prey population and $\mathcal{K}$ the 
carrying capacity of the prey genotypes. The logistic term involves exponential growth for small population numbers and saturation as population values approach the carrying capacity.
Moreover $1/A_i$ is the yield coefficient 
of prey genotype $i$ to predator genotype $i$. Equation \eqref{ht} is a Holling ``type II'' functional response 
\citep{Case2000,Hastings1991}, where predation rate is a saturating function 
of prey density. $C_i$ and $k^p$ are constants parametrizing the saturating 
functional response. The constant $k^p$, which describes the maximal 
predation rate, can also be interpreted as the predator's maximal self-replication 
or intrinsic growth rate.

Both terms $\sum_{<j>_i} x_j - x_i$, and $\sum_{<j>_i} y_j - y_i$, denote genetic 
diffusion by mutation among neighboring genotypes for both prey and 
predator genotypes, which are proportional to $\mu^h$ and $\mu^p$, 
respectively. Moreover, $k^h_i$ denotes self-replication constant 
(intrinsic growth rates) for prey genotypes; $\epsilon^h_i$ and $\epsilon^p_i$ are 
decay rates which can be interpreted as spontaneous hydrolysis rates as well 
as density-independent death rates. If only a single genotype is present in each 
species, equations \eqref{g_RQD1} and \eqref{g_RQD2} are close to the well-known Rosenzweig-MacArthur model \citep{Weltz2006}. 

As the reader will see, since the dimension of the dynamical system described by Eqs. \eqref{g_RQD1}-\eqref{g_RQD2} depends upon the length of the sequences, one may expect different types of dynamics for different values of $\nu$. As we will discuss in the following sections, were we review results for $\nu =1$ \cite{Sardanyes2008a} (figure \ref{figure10}(a)) and $\nu = 3$ \cite{Sardanyes2007b} (figure \ref{figure8}(c)), this is the case. The fitness landscape for this predator-prey system using $\nu =1$ is shown in figure \ref{figure10}(a). Actually, this system was studied for two different fitness landscapes, given by a flat or neutral fitness landscape, with $k^h_i \equiv k^h$ and $k^p_i \equiv k^p$, $\forall i$ (i.e., all genotypes share the same fitness values) and for an asymmetric fitness landscape with $k^p < K^p$ or $k^p > K^p$ (i.e., one of predator's genotypes has a higher fitness in terms of populations growth). For this latter case, a predator with a higher fitness actually means that a given genotype is more efficient in catching its preferred prey. For both scenarios we also assumed that mutation rates for both prey and predator genotypes were equal 
i.e. $\mu^h_i \equiv \mu^h$ and $\mu^p_i \equiv \mu^p$, $\forall i$ (see \cite{Sardanyes2008a} for further details). 

The analysis of the qualitative behavior of Eqs. \eqref{g_RQD1}-\eqref{g_RQD2} was performed assuming a neutral fitness landscape with $\epsilon^h_i = \epsilon^p_i \equiv \epsilon$. This system was shown to have three fixed points, given by: $(x^*_i=0,y^*_i=0)$, $(x^*_0,x^*_1,0,0)$, and $(x^*_0,x^*_1,y^*_0,y^*_1)$. The first fixed point, if stable, involved predator-prey extinction. The second equilibrium, involved prey survival and predator's extinction whereas the third fixed point involved predator-prey coexistence, and thus it was the potential state of Red Queen dynamics. This third fixed point, under symmetry conditions, named $({\bf{x^*}},{\bf{y^*}})$, was given by:
\begin{equation}
\nonumber{
x^* = \frac{\epsilon^p}{k^p-\epsilon^p}},
\end{equation}
\begin{equation}
\nonumber{
y^* = \frac{(1+x^*)(k^h - 2k^h x^* - \epsilon^h)}{k^p}}.
\end{equation}
To study the stability of this fixed point it was further assumed that $k^h = k^p = 1$, also considering symmetry in decay rates i.e., $\epsilon^h = \epsilon^p = \epsilon$. Under this conditions, the fixed point reads:
\begin{equation}
\nonumber{
x^* = \frac{\epsilon}{1-\epsilon}},
\end{equation}
\begin{equation}
\nonumber{
y^* = \frac{(1 - 4 \epsilon + \epsilon^2)}{(1-\epsilon)^2}}.
\end{equation}
After some algebra, and after fixing $\mu^h = \mu^p \equiv \mu$, a critical mutation rate causing a Hopf bifurcation was identified at:
\begin{equation}
\nonumber{
\mu_c =  \frac{\epsilon(1-4 \epsilon +  \epsilon^2)}{4(1-\epsilon)}}. 
\end{equation}

Such a bifurcation, which involves the creation of a periodic orbit causing sustained, periodic oscillations, was confirmed by numerical simulations (figure \ref{figure10}(b)). Interestingly, the same bifurcation was also numerically found for the fitness landscape with asymmetries in predator's replication rates. Counterintuitively, the asymmetric fitness landscape revealed that the most efficient predator genotypes achieved lower population equilibria (see \cite{Sardanyes2008a} for details). Our results identifying periodic Red Queen dynamics were in agreement with other mathematical models on coevolution (see \cite{Dieckmann1995,Dercole2012}).

\subsection{Chaotic Red Queen attractors}

So far, we have discussed the dynamical behavior of Eq. \eqref{g_RQD1}-\eqref{g_RQD2} for $\nu = 1$, which can be governed by stable fixed points as well as by a periodic orbit causing sustained and regular oscillations of predator-prey genotypes populations. The same model, analyzed for $\nu = 3$, revealed much richer dynamics: under some parameter regions, both populations behave chaotically. Hence, similarly to what is known as diffusion-induced chaos \cite{Pascual1993}, it was found that the simplest system (with $\nu =1$), governed by a periodic orbit, could be governed by chaotic attractors at increasing the number of available alleles (more available nodes in sequences space). 

It is known that dynamical systems governed by a periodic orbit can become unstabilized to chaos when spatial correlations and diffusion are included. Actually, some decades ago a great deal of attention was paid to self-organization processes in reaction-diffusion systems, and their relevance in chemistry, physics and biology was repeatedly stressed \cite{Nicolis1977,Haken1983,Vidal1984}.  In this sense, numerical investigations of the spatially-extended Belousov-Zhabotinsky chemical reaction showed the presence of chaotically oscillating structures. Moreover, diffusion-induced chaos has also been discussed in the context of spatial ecological dynamics \cite{Pascual1993}. We notice that MA chaotic dynamics can be also interpreted from the perspective of dynamical unstabilization due to diffusion in space. That is, each sequence of the sequences space can be interpreted as a patch, and populations can diffuse between patches because of mutation (i.e., diffusion in sequences space). Under this view, the oscillatory behavior of variables $x_i$ and $y_i$ in Eqs. \eqref{g_RQD1}-\eqref{g_RQD2} for $\nu = 1$, becomes unstabilized to chaos for $\nu=3$.
\begin{figure*}
\center
\includegraphics[width=0.8\textwidth]{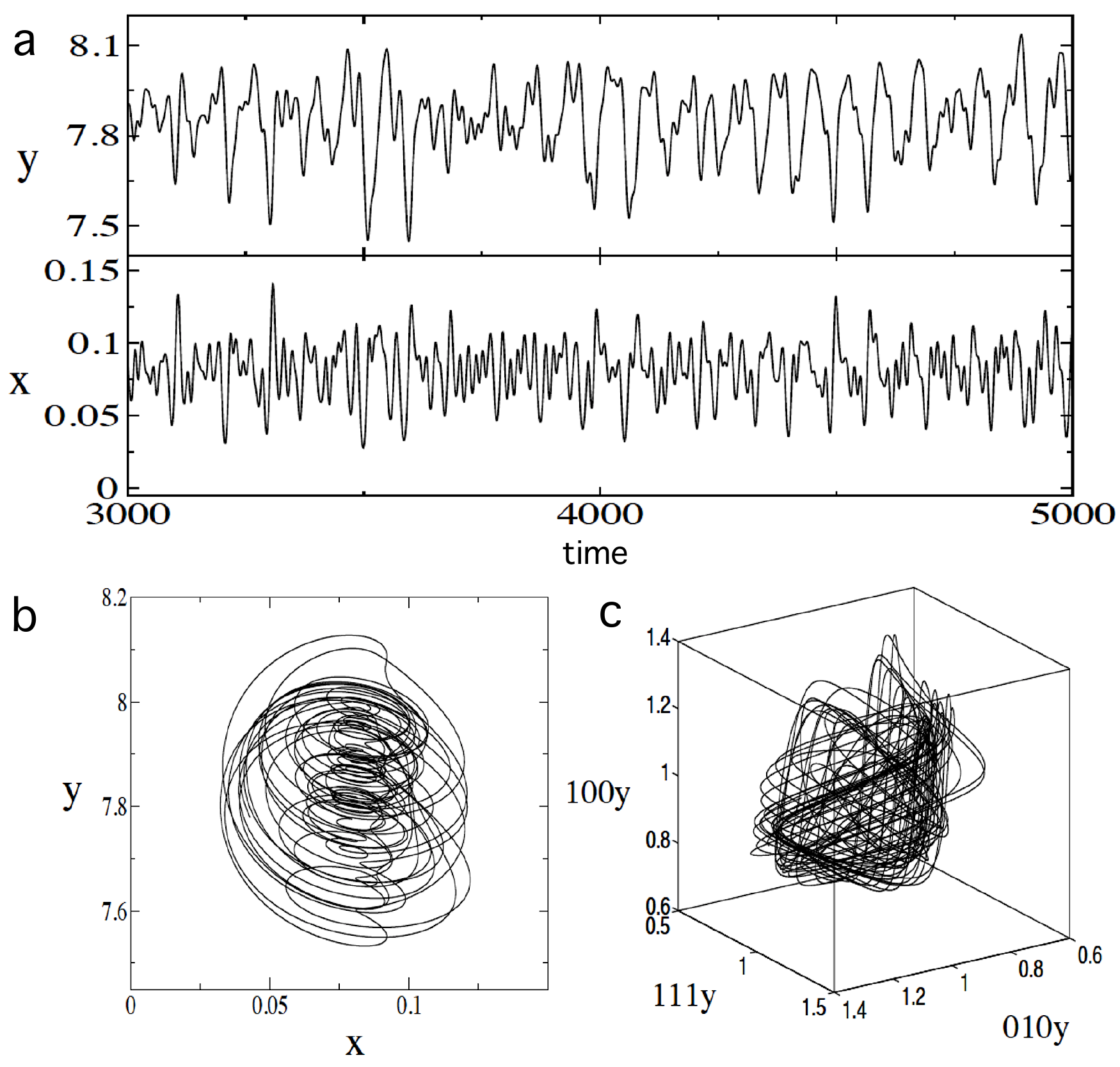}
\caption{Red Queen chaotic dynamics for Eqs. \eqref{g_RQD1}-\eqref{g_RQD2} using $\nu = 3$. (a) Global population time dynamics of parasites ($y$) and host ($x$). (b) Strange attractor governing host-parasite global dynamics. In (c) we show the chaotic attractor in the parasites three-dimensional phase space for genotypes $(010y,111y,100y)$ (see \cite{Sardanyes2007b}).}
\label{figure11}
\end{figure*}

Figure \ref{figure11} shows the chaotic coevolutionary dynamics for the $3$-bits sequences modeled with Eqs. \eqref{g_RQD1}-\eqref{g_RQD2}. The time series in figure \ref{figure11}(a) are represented for the global populations of predators ($y$) and preys ($x$). In figure \ref{figure11}(a) and (b) we show, respectively, the chaotic attractors for global host-parasite populations (represented in the phase space $(x,y)$) as well as the attractor for parasite genotypes $010$, $111$, and $100$ (see \cite{Sardanyes2007b} for further details). 

Our previous results suggested that Red Queen dynamics can be chaotic even for small haploid replicator systems with MA interactions. It was previously shown that large networks with host-parasitoid replicators can also behave chaotically. More specifically, Kaneko and Ikegami \cite{Ikegami1992,Kaneko1992} characterized the so-called homeochaos in multi-species models with antagonistic interactions and evolution. They suggested that chaos, more than a destabilizing behavior \cite{Berryman1986}, could involve stability in multi-species ecosystems through a weak, chaotic state arising in high-dimensional dynamical systems. Roughly, homeochaos was suggested to suppress strong chaos causing large fluctuations that could near populations to extinction. Homeochaos is characterized by many positive, but close to zero Lyapunov exponents (i.e., a type of hyperchaos). Such a property of the spectrum of Lyapunov exponents involves narrow chaotic fluctuations with small amplitude, which are able to keep population numbers far away from attractors involving extinction. The concept of homeochaos was later extended to low-dimensional systems, and its role was discussed in both deterministic and stochastic host-parasitoid models with discrete time generations \cite{Sardanyes2011b}.

Chaotic evolutionary dynamics have been found in other theoretical studies of genetic polymorphisms under frequency-dependent selection (see for example \cite{May1983,Seger1992,Ferriere1995}). Moreover, Dercole and colleagues \cite{Dercole2012} recently showed that predator-prey coevolutionary models governed by periodic fluctuations became chaotic when the system is embedded in a three-species food chain model by the addition of a superpredator able to coevolve. These authors argued that over space, genetically-driven chaos may cause evolutionary divergence of local metapopulations, even under the absence of environmental change, thus promoting genetic diversity among ecological communities over long evolutionary time.

\section{Large Scale Coevolution on Complex Networks}

Our last example deals with the large-scale evolutionary dynamics. The study of large scale coevolution was also performed in multi-species models using complex networks theory. 
A very simple model of large-scale evolution involving a set of $N$
interacting species can be easily defined \cite{Sole1996,Sole1996a,Sole1996b,Sole1997}. In this model, species 
interactions are introduced by means of a $N \times N$ connectivity matrix ${\bf
W}=(W_{ij})$. Evolution for this system is introduced through changes in its elements. Similarly to some of the models previously presented, here the "state" of each species is described by
a binary variable $S_i \in \{0, 1\}$ ($i=1,2,...,N$), 
for the $i$-th species, with $S_i=1$ or $S_i=0$ if the species
is alive or extinct, respectively. So the whole ecosystem is described in terms of 
a simple directed graph where the connections are initially set to random values.
Each species receives either positive or negative inputs from other species. These signs indicate 
that the given species is favored or harmed by the species which send the input. For instance, a negative input would correspond to the interaction with a predator or with a parasite. Alternatively, a positive input would correspond to mutualism or symbiosis. Such a model, in its
simplest form, can be formulated in terms of a set of rules displayed in Box 3.

\begin{figure*}
\center
\includegraphics[width=1.0\textwidth]{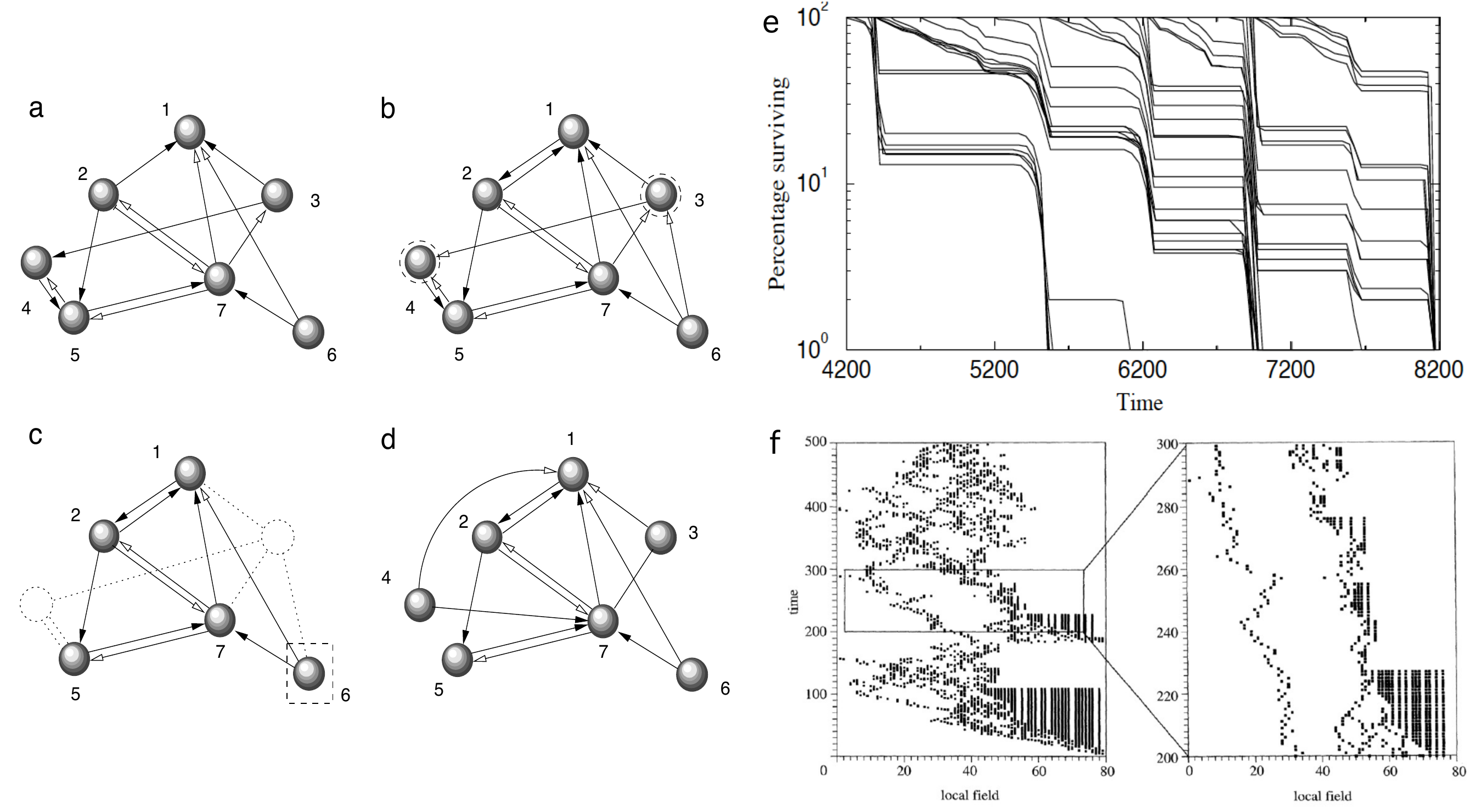}
\caption{(a-d) Rules of the evolution model exemplified with a network with $N=7$ nodes and a given initial connectivity (a). Negative and positive interactions are indicated with black and white arrows, respectively. The first step of the model is to modify the links (b): here different pairs of interactions are found, including mutualism, parasitism, predation and competition. For this particular case, species $3$ and $4$ become nonviable, dying in the next step (c). Species $6$ is selected (among the survivors) and copied into the two empty sites (Rule 3). Such copies carry the links of their parent species, which will be modified again by the first rule in an iterative process. 
This set of rules generates a very complex dynamical pattern of species evolution.  In (e) we display an example of the 
survival curves obtained from our model (compare with figure \ref{figure1}). In (f) we also show the evolution of the local fields over time. A mixture of slow and rapid changes occur, in a punctuated fashion (see \cite{Sole1996a} for further details).}
\label{figure12}
\end{figure*}

\begin{svgraybox}
{\bf{Box 3.}} Rules of the network coevolution model (illustrated in figure \ref{figure12}(a-d)):  
 
\begin{enumerate}
\item
{\bf{Random changes in the connectivity matrix}}. At each generation, we select one input connection for each species and assign it to a new, random value without regard for the previous state of the connection. This rule introduces changes into
the web, which can be due to evolutionary responses 
or to environmental changes of some sort. In other words, 
changes derived from coevolution among two species, innovation at the
 species level and/or environmental-driven changes are lumped together
 within this rule.\\

\item
{\bf{Extinction}}. Changes in the connectivity will eventually lead to extinctions. Extinction events are decided by computing the total sum of the inputs for each species. This sum, if negative will involve the extinction of the species ($S_i=0$), and all its connections are removed. 
Otherwise, nothing happens ($S_i=1$). Hence, the state of the $i$-th species is updated following the following dynamical equation:
\begin{equation}
\nonumber{
S_i(t+1) = \Phi \Biggl [\sum_{j=1}^N W(i,j) \Biggr ]}
\end{equation} 
where  $\Phi(z)=1$ for positive $z$ and zero otherwise.\\

\item
{\bf{Diversification}}. A number of species can disappear due to the extinction rule, leaving empty sites. These sites will be refilled by diversification: each extinct species is replaced by a randomly chosen survivor. The replacement is made by simply copying the connections of the
survivor into the empty site.

\end{enumerate}
\end{svgraybox}

This model shows a strongly
nonlinear behavior with avalanches of extinction as well as the correct
power law distribution of extinction sizes \cite{Sole1996,Sole1996b,Manrubia1998,Sole1998} (although with an exponent typically close but higher
than $\alpha=2$, see also \cite{Drossel2001}). The outcome of the model was that both small and very large events were generated {\em by the same} dynamical
rules. Most of the times, the extinction of a given species had
no consequences for the other species. But from time to time, a given 
(keystone) species with positive inputs to others disappeared. 
The removal of this species was suggested to have a destabilizing effect on
others, able to cause further propagation leading to mass extinction events. 

What is the origin of such extinction patterns? We first need to see 
how a given species can shift towards a negative sum of inputs. The
reason is easily understandable from rule 1 above. 

Since changes of links
among species are random and the new values are chosen from a uniform
distribution an expected consequence is that, in the long run, the sum
of inputs will decay to zero. If we look at the sign of the links, so
that the probability of finding positive links, $P(W^+)=P[W_{ij}>0]$; and the probability of finding negative links $P(W^-)=P[W_{ij}<0]$. The time
evolution of the positive links can be described in terms of a master
equation, given by:

\begin{equation}
\nonumber{
{dP(W^+,t) \over dt} = w(W^-\rightarrow W^+)P(W^+)-
w(W^+\rightarrow W^-)P(W^-)},
\end{equation}
where $P(W^+)+P(W^-)=1$, starting for example from an
initial condition $P(W^+,0)=P_0$. Since, from the first rule, we have 
$w(W^-\rightarrow W^+)=w(W^+\rightarrow W^-)=1/2N$, the master equation
reads:
\begin{equation}
\nonumber{
{dP(W^+,t) \over dt} = {1 \over 2N} \left [ 1 - P(W^+) \right ]},
\end{equation}
and an exponential decay is obtained:
\begin{equation}
\nonumber{
P(W^+,t) = {1 \over 2} \left [ 1 + (2P_0-1) e^{-t/N} \right ]}.
\end{equation}

As a consequence, the sum of inputs ${\cal F}_i=\sum_j W_{ij}$ will
also decay exponentially: ${\cal F}_i(t) \sim e^{-t/N}$, predicting 
an exponential decay in the probabilities of
survival, as expected from the Red Queen hypothesis (see Introduction). This rule
actually introduces the basic ingredients of Van Valen's theory. All
species in the system keep changing all the time (either due to
biological or environmental causes) eventually reaching extinction. 
The ultimate fate of all species is to get extinct, and so an 
exponential decay in the survival probability will be observed. Here,
however, there is no intrinsic, species-level variability (in terms of
a genotype) and the fate
of a given species will be dominated by network responses and
chance. Ecological-driven phenomena are the key forces in the long run,
although small-scale events are taking place all the time. The nature
of the decay turns to be exponential on average but episodic
when looking at pseudocohorts (see figure \ref{figure1}), consistently with Raup's
analysis commented on in the Introduction of this chapter (compare figures \ref{figure1} and \ref{figure12}).

An analytic study of the previous model is rather difficult because of the random nature of the interaction matrix. The model can, however, be simplified by mapping the set of rules into a linear model \cite{Manrubia1998,Sole1998}, giving place to a mean field approach to the
network model (see Box 4). 

\begin{svgraybox}
{\bf{Box 4.}} Consider a set of $N$ species, characterized
by a single integer quantity $\phi_i$ ($i=1,2,...,N)$. This quantity
will play the role of the internal field. Each species is now represented 
by this single
(integer) number $\phi_i \in \{-N, -N+1, ..., -1, 0, 1, ..., N-1, N\}$, 
which represents the sum of inputs from other species. The dynamics consists of three steps: (a) with probability $P=1/2$,
$\phi_i \rightarrow \phi_i-1$, otherwise no change occurs (this is equivalent
to the randomization rule in the network model); (b) all species
with $\phi_i < \phi_c$ (below a given threshold) are extinct. 
Here we use $\phi_c=0$ but other choices give the same results. The number of
extinct species, $0 < E < N$, gives the size of the extinction event. All $E$ extinct
species are replaced by survivors. Specifically, for each extinct site
(i. e. when $\phi_j < \phi_c$) we choose one of the $N-E$ survivors $\phi_k$ and
update $\phi_j$ to: $\phi_j = \phi_k$; (c) after an extinction event, a wide reorganization
of the web structure occurs \cite{Sole1997}. In this simplified
model this is introduced as a coherent shock. Each of the survivors are
updated as $\phi_k=\phi_k+q(E)$, where $q(E)$ is a random integer
between $-E$ and $+E$. This mean-field approach defines a three-step process. If $N(\phi)$ indicates the frequency of species having a local field $\phi$, we have:
\begin{equation}
\nonumber{
N(\phi, t+1/3) = {1 \over 2} N(\phi,t) + {1 \over 2} N(\phi+1,t)},
\end{equation}
\begin{equation}
\nonumber{
N(\phi, t+2/3) =  N(\phi,t+1/3) + N(\phi,t+1/3) \sum_m {m \over N-m} P(m)},
\end{equation}
if $\phi>0$ and zero otherwise. Finally:
\begin{equation}
\nonumber{
N(\phi, t+1) =  N(\phi,t+2/3) - N(\phi,t+1/3) + \sum_{q > -\phi}
N(\phi+q, t+1/3) P(q)},
\end{equation}
from these equations, the full master equation for the dynamics reads:
$$N(\phi, t+1) =  {1 \over 2}
\sum_{q = - \infty}^{+ \infty} \sum_m 
{P(m) \over 2m+1} \theta( m - \vert q \vert) 
\Biggl [
N(\phi+q, t)- N(\phi+q+1,t) \Biggr ]$$
\begin{equation}
\nonumber{
 + {1 \over 2} 
[N(\phi,t) + N(\phi+1,t)] \sum_m {m P(m) \over N - m}}.
\end{equation}
Where two basic statistical distributions, which are self-consistently related,
have been used. These are:
\begin{equation}
\nonumber{
P^*(q) = \sum_m {P_e(m) \over 2m +1 } \theta (m - \vert q \vert )},
\end{equation}
which is an exact equation giving the probability of having a shock of
size $q$. The second is $P_e(m)$, the extinction probability for an event
of size $m$. We have a mean-field approximation relating both distributions:
$$ P_e(m) = \sum_q P^* (q) \delta 
\Bigl [ \sum_{\phi=1}^{q-1} N(\phi) - m \Bigr ]. \eqno$$
The last equation introduces the average profile $N(\phi)$, i. e. the 
time-averaged distribution of $\phi$-values. For the mesoscopic
regime $1 \gg q \gg N$, by applying a Taylor expansion to the master equation, i. e. 
$$ N(\phi) = {1 \over 2}
\sum_{q = - \infty}^{+ \infty} \sum_m 
{P(m) \over 2m+1} 
\Biggl \{
2 N(\phi+q) + 
{\partial N \over \partial \phi} \Biggl \vert_{\phi + q} +
{1 \over 2} {\partial^2 N \over \partial \phi^2}\Biggl \vert_{\phi + q} + ...
\Biggr \} + $$
$$ + {1 \over 2}
\Biggl \{
2 N(\phi) + 
{\partial N \over \partial \phi} \Biggl\vert_{\phi} +
{1 \over 2} {\partial^2 N \over \partial \phi^2} \Biggl\vert_{\phi} + ...
\Biggr \}
 \sum_m {m P(m) \over N - m},$$ 
and using a continuous approximation, it is easy to see that the
previous equation reads:
$${1 \over 2} \int dm \int_{-m}^{m} {P(m) \over 2 m}
\Biggl \{
2  \Bigl [
{ N(\phi+q) \over N(\phi)} - 1 
\Bigr ] +
{\partial Ln N \over \partial \phi} \Biggl\vert_{\phi+q} + ...
\Biggr \} + $$
$$+ {1 \over 2} 
\Biggl \{
2 + {\partial Ln N \over \partial \phi}\Biggl \vert_{\phi} + ...
\Biggr \} \int {m P(m) \over N-m } dm = 0.$$
Assuming that $N(\phi)$ decays exponentially,
i. e. $N(\phi)=\exp(-c\phi/N)$, 
we can integrate each part of the last equation,
using $N(\phi + q)/N(\phi) = \exp(-cq/N) \approx 1-cq/N$.
The first term cancels exactly, the second gives
$-2c/N$ and the third scales as $(1-O(1/N))N^{1-\tau}$. So the
previous equation leads to:
\begin{equation}
\nonumber{
- {2 c \over N} + N^{1-\tau} G \Bigl 
[ 1 - O \left ({1 \over N} \right ) \Bigr ]  = 0},
\end{equation}
in order to satisfy this equality, we have $\tau=2$, which gives us
the scaling exponent for the extinction distribution. Hence, in agreement
with Burlando's analysis, the taxonomy that emerges from this model also displays
fractal behavior (with an exponent $\alpha_b \approx 2$). 
\end{svgraybox}

These models, able to reproduce observed patterns, can 
have important implications for evolutionary theory. An intense debate over the last decades 
has concerned the basic mechanisms operating at different temporal scales. Some authors (specially in the field of population genetics) suggested that the rules operating at small scales (i.e., microevolutionary events) can be directly translated into the process of 
macroevolution \cite{Haldane1932,Dobzhanski1951,Li1991}. 
However others authors like Stephen Jay Gould, claimed that different processes 
are at work in evolution at different scales \cite{Gould2003}, although no well-defined mechanism for such decoupling was proposed. 
We want to notice that the network organization of ecologies, changing in a coevolving landscape, suggests a possible source of decoupling the micro- and macro scales. Moreover, these models can also help understanding the complex dynamical behavior of large extinctions and their aftermath \cite{Sole2002,Sole2010,Roopnarien2006,Benton2012}.

\section{Conclusions}

Coevolutionary dynamics introduces an additional layer beyond single-species 
evolution. Coevolution pervades biology on multiple scales but its role and impact 
-as illustrated by our previous examples- is rather different at each scale. In microorganisms, 
the changes that couple diverse species (as hosts and pathogens) within a given 
ecosystem are associated to many different molecular events related to membrane receptors, 
production and pumping of toxins, development of aggregates or resistance to antibiotics, to cite just a few. 
Arm races are known to occur and take place on short time scales. 

Coevolution occurs in other systems, including technological ones. Coevolution between 
predators and prey predate at least part of the evolutionary events that triggered the emergence 
of complex animals at the base of the Cambrian explosion. It is likely that the so called Ediacaran 
fauna, dominated by simple, filtering organisms with small developmental 
complexity became replaced by the well known, Burguess-Shale pattern as a 
consequence of predator-prey arm races.  Many challenges lie ahead in our understanding of how 
coevolution shaped biological complexity and how to properly approach it from a theoretical 
perspective. Among other questions, we still need to understand how to 
connect ecological networks and coevolving landscapes, how to place these landscapes in the middle 
of the multidimensional space involving development, ecology and the environment, and what 
universal trends are to be found in their structure and dynamics. The previous examples only provide a glimpse 
of the richness and complexity, but they also illustrate the power of simple models able to address 
relevant questions.

\begin{acknowledgement}
We want to thank Sergi Valverde for help in model implementation 
and the other members of the Complex Systems 
Lab for useful discussions. This work has been funded by the 
Bot\'in Foundation, by the James S. McDonnell foundation and by the Santa Fe Institute.\\

{\bf{The present document is a chapter of the book titled \emph{Recent Advances in Theory and Application of Fitness Landscapes} (Published by Springer). The final publication is available at http://link.springer.com/}}

\end{acknowledgement}
%
%%%%%%%%%%%%%%%%%%%%%%%% referenc.tex %%%%%%%%%%%%%%%%%%%%%%%%%%%%%%
% sample references
% %
% Use this file as a template for your own input.
%
%%%%%%%%%%%%%%%%%%%%%%%% Springer-Verlag %%%%%%%%%%%%%%%%%%%%%%%%%%
%
% BibTeX users please use
% \bibliographystyle{}
% \bibliography{}

\begin{thebibliography}{99.}%
% Use the following syntax and markup for your references if 
% the subject of your book is from the field 
% "Mathematics, Physics, Statistics, Computer Science"
%
% Contribution 

\bibitem{Agrawal2002}
Agrawal, A., Lively, C. M. Infection genetics: gene-for-gene versus matching-alleles models and all points in between. Evol. Ecol. Res. 4, 79-90 (2002)

\bibitem{Agrawal2001}
Agrawal, A. F., Lively, C. M. Parasites and the evolution of self-fertilization. Evolution 55(5), 869-879 (2001)

\bibitem{Bak1992}
Bak, P., Flyvjerg, H., Lautrup, B. Coevolution in a Rugged Fitness Landscape. Phys. Rev. A 46, 6724-6730 (1992)

\bibitem{Benton1995}
Benton, M. J. \emph{Red Queen hypothesis}. In: Paleobiology, eds. D.E.G. Briggs and P.R. Growther, Blackwell: Oxford (1995)

\bibitem{Berryman1986}
Berryman, A. A., Millstein, J. A. Are ecological systems chaotic - and if not, why not? Trends Ecol. Evol. 4, 17-28 (1986)

\bibitem{Case2000}
Case, T.J. An illustrated guide to theoretical ecology. Predator-prey systems: predator dynamics and effects on prey. Oxford (New York). Oxford University Press (2000)

\bibitem{Benton2012}
Chen, Z-Q., Benton, M. J. The timing and pattern of biotic recovery following the end-Permian mass extinction. Nat. Geosci. 5, 375-383 (2012)

\bibitem{Clarke1994}
Clarke, D. K., Duarte, E. A., Elena, S. F., Moya, A., Domingo, E., Holland, J. J. The Red Queen reigns in the kingdom of RNA viruses. Proc. Natl. Acad. Sci. U.S.A. 91, 4821-4824 (1994)

\bibitem{Darwin1859}
Darwin, C. \emph{On the Origin of Species by Means of Natural Selection, or the Preservation of Favoured Races in the Struggle for Life}, London: John Murray (1859)

\bibitem{Decaestecker2007}
Decaestecker, E., Gaba, S., Raeymaekers, J. A. M., Stoks, R., Van Kerckhoven, Ebert, D., Meester, L. D. Host-parasite 'Red Queen'dynamics archived in pond sediment. Nature 450, 870-873 (2007)

\bibitem{Dercole2012}
Dercole, F., Ferriere, R., Rinaldi, S. Chaotic Red Queen coevolution in a three species food chain. Proc. Roy. Soc. B 277, 2321-2330 (2012)

\bibitem{deVisser2007}
de Visser, J. A. G. M., Elena, S. F. The evolution of sex: empirical insights into the roles of epistasis and drift. Nat. Rev. Genetics 8, 139-149 (2007)

\bibitem{Dieckmann1995}
Dieckmann, U., Marrow, P., Law, R. Evolutionary cycling in predator-prey interactions: Population dynamics and the Red Queen. J. Theor. Biol. 176, 91-92 (1995)

\bibitem{Dobzhanski1951}
Dobzhansky, T. \emph{Genetics and the origin of species} (3rd edition). Columbia University Press, New York (1951)

\bibitem{Drossel2001}
Drossel, B. Biological evolution and statistical physics. Adv. Phys. 50, 209-295 (2001)

\bibitem{Ehrlich1964}
Ehrlich, P.R., Raven, P.H. Butterflies and plants: A study in coevolution. Evolution 18, 586-608 (1964)

\bibitem{Eigen1971}
Eigen, M. Selforganization of matter and evolution of biological macromolecules. Naturwiss. 58, 465-523 (1971)

\bibitem{Elena2010}
Elena, S. F., Sol\'e, R. V., Sardany\'es, J. Simple genomes, complex interactions: Epistasis in RNA virus. Chaos 20, 026106 (2010)

\bibitem{Ferriere1995}
Ferri\`ere, R. Fox, G. A. Chaos and evolution. Trends Ecol. Evol. 10, 480-485 (1995)

\bibitem{Flor1956}
Flor, HH.: The complementary genetic systems in flax and flax rust. Adv. Genetics 8, 29-54 (1956)

\bibitem{Freund1991}
Freund, H., Wolter, R. Evolution of bit strings: some preliminary results. Complex Systems 5, 279-298 (1991)

\bibitem{Gould2003}
Gould, S. J. \emph{The structure of evolutionary theory}. Harvard University Press, Cambrigde, MA (2003)

\bibitem{Grosberg2000}
Grosberg, R.K. Hart, M.W. Mate selection and the evolution of highly polymorphic self/nonself recognition genes. Science 289, 2111-2114 (2000)

\bibitem{Haken1983}
Haken, H. \emph{Advanced Synergetics}. Springer Series in Synergetics. Springer-Verlag: New York (1983)

\bibitem{Haldane1932}
Haldane, J. B. S. \emph{The causes of evolution}. Longmans and Green, London (1932)

\bibitem{Hamilton1980}
Hamilton, W. D. Sex vs. non-sex vs. parasite. Oikos 35, 282-290 (1980)

\bibitem{Hamilton1990}
Hamilton,  W.D., Axelrod, A., Tanese, R. Sexual reproduction as an adaptation to resist parasites (a review). Proc. Natl. Acad. Sci. U.S.A. 87, 3566-3573 (1990)

\bibitem{Hastings1991}
Hastings, A., Powell, T. Chaos in a three-species food chain. Ecology 72(3), 896-903 (1991)

\bibitem{Hoffman1991}
Hoffman, A. Testing the Red Queen hypothesis. J. Evol. Biol. 4, 1-7 (1991)

\bibitem{Howard1994}
Howard, R. S., Lively, C.M. Parasitism, mutation accumulation and the maintenance of sex. Nature 367, 554-557 (1994)

\bibitem{Ikegami1992}
Ikegami, T., Kaneko, K. Evolution of host-parasitoid network through homeochaotic dynamics. Chaos 2, 397-407 (1992)

\bibitem{Ilachinsky2000}
Ilachinsky, A. \emph{Cellular Automata. A Discrete Universe}. World Scientific: Singapore (2000)

\bibitem{Jacob1977}
Jacob, F. Evolution and tinkering. Science 196, 1161 (1977)

\bibitem{Jacob1983}
Jacob, F. Molecular tinkering in evolution. In \emph{Evolution from Molecules to Men} (D.S. Rondall, ed.). Cambridge University Press, Cambridge (1983)

\bibitem{Jaenike1978}
Jaenike, J. An hypothesis to account for the maintenance of sex in populations. Evol. Theor. 3, 191-194 (1978)

\bibitem{Kaneko1992}
Kaneko, K., Ikegami, T. Homeochaos: dynamic stability of a symbiotic network with population dynamics and evolving mutation rates. Physica D 56, 406-429 (1992)

\bibitem{Kauffman1991}
Kauffman, S. A., Johnsen, J. Coevolution on the edge of chaos: Coupled fitness landscapes, poised states and coevolutionary avalanches. J. Theor. Biol. 149, 467-505 (1991)

\bibitem {Kauffman1993}
Kauffman, S. A. \emph{The Origins of Order}. Oxford University Press: New York (1993)

\bibitem{Kerr1987}
Kerr, A. The impact of molecular genetics of plant pathology. Annu. Rev. Phytopathol. 25, 87-110 (1987)

\bibitem{King2009}
King, K. C., Delph, L. F., Jokela, J., Lively, C. M. The geographic mosaic of sex and the Red Queen. Curr. Biol. 19, 1438-1441 (2009)

\bibitem{Lafforgue2011}
Lafforgue, G., Mart\'inez, F., Sardany\'es, J., de la Iglesia, F., Shi-Shun, L., Qi-Wen, N., Sol\'e, R. V., Chua, N. H., Dar\'os, J-A., Elena, S. F. Tempo and mode of plant RNA virus escape from RNA interference-mediated resistance. J. Virol. 85(19), 9686-9695 (2011)

\bibitem{Li1991}
Li, W. H., Graur, D. \emph{Fundamentals of molecular evolution}. Sinauer Associates, Sunderland, MA (1991)

\bibitem{Lively1987}
Lively, C. M. Evidence from a New Zealand snail for the maintenance of sex by parasitism. Nature 328, 519-521 (1987)

\bibitem{Manrubia1998}
Manrubia, S. C., Paczuski, M. A simple model of large-scale organization in Evolution. Int. J. Mod. Phys. C 9, 1025-1032 (1998)

\bibitem{May1983}
May, R. M. Anderson, R. M. Epidemiology and genetics in the coevolution of parasites and hosts. Proc. R. Soc. Lond. B 219, 281- 313 (1983)

\bibitem{McCaskill2001}
McCaskill, J.S. and Altemeyer, S.,  Error threshold for spatially resolved evolution in the quasispecies model. Phys. Rev. Lett. 86, 5819 (2001)

\bibitem{Mode1958}
Mode, D.J.  A mathematical model for the co-evolution of obligate parasites and their hosts. Evolution 12, 158-165 (1958)

\bibitem{Montoya2006}
Montoya, J.M., Pim, S., Sol\'e, R.V. Ecological networks and their fragility. Nature 442, 259-264 (2006)

\bibitem{Morran2011}
Morran, L. T., Schmidt, O. G., Gelarden, I. A., Parrish II, R. C., Lively, C. M. Running with the Red Queen: Host-parasite coevolution selects for biparental sex. Science 333, 216-218 (2011)

\bibitem{Newman2003}
Newman, M. E. J., Palmer, R. G. \emph{Modeling Extinction}: Oxford University Press, New York.

\bibitem{Nicolis1977}
Nicolis, G., Prigogine, I. \emph{Self-Organization in Non-Equilibrium Systems}. Wiley-Interscience: New York, 1977

\bibitem{Parker1994}
Parker, M. A. Pathogens and sex in plants. Evol. Ecol. 8, 560-584 (1994)

\bibitem{Pascual1993}
Pascual, M. Diffusion-induced chaos in a spatial predator-prey system. Proc. Roy. Soc. London B 251, 1-7 (1993)

\bibitem{Perelson1991}
Perelson, A.S., Kauffman, S. (eds). \emph{Molecular evolution on rugged landscapes: proteins, RNA and the immune system}. In: SFI Studies in the Sciences of Complexity Vol. IX. Redwood, CA: Addison-Wesley (1991)

\bibitem{Quer1996}
Quer, J., Huerta, R., Novella, I. S., Tsimring, L., Domingo, E., Holland, J. J. Reproducible nonlinear population dynamics and critical points during replicative competitions of RNA virus quasispecies. J. Mol. Biol. 264, 465-471 (1996)

\bibitem{Raup1986}
Raup, D.M. Biological extinction and Earth history. Science 231, 1528-1533 (1986)

\bibitem{Raup1991}
Raup, D. M. A kill curve for phanerozoic marine species. Paleobiology 17, 37-48 (1991)

\bibitem{Roopnarien2006}
Roopnarine, P.D., Extinction cascades and catastrophe in ancient food webs. Paleobiology 32 (1), 1 (2006)

\bibitem {Sardanyes2007a}
Sardany\'es, J., Sol\'e, R. V. Chaotic stability in spatially-resovled host-parasite replicators: The Red Queen on a lattice. Int. J. Bif. and Chaos 17(2), 589-606 (2007)

\bibitem {Sardanyes2011b}
Sardany\'es, J. Low dimensional homeochaos in coevolving host-parasotoid dimorphic populations: Extinction thresholds under local noise. Commun. Nonlinear Sci. Numer. Simul. 16, 3896-3903 (2011)

\bibitem {Sardanyes2008a}
Sardany\'es, J., Sol\'e, R. V. Matching allele dynamics and coevolution in a minimal predator-prey replicator model. Phys. Lett. A 372, 341-350 (2008)

\bibitem {Sardanyes2007b}
Sardany\'es, J., Sol\'e, R. V. Red Queen strange attractors in host-parasite replicator gene-for-gene coevolution. Chaos, Solitons and Fractals 32(5), 1666-1678 (2007)

\bibitem{Sardanyes2010}
Sardany\'es, J., Elena, S. F. Error threshold in RNA quasispecies models with complementation. J. theor. Biol. 265(3), 278-286 (2010)

\bibitem{Sardanyes2011a}
Sardany\'es, J., Elena, S.F. Quasispecies spatial models for RNA viruses with different replication modes and infection strategies. PLoS ONE 6(9), e24884 (2011)

\bibitem{Sardanyes2009}
Sardany\'es, J., Sol\'e, R. V., Elena, S. F. Replication mode and landscape topology differentially affect RNA virus mutational load and robustness. J. Virol. 83(23), 12579-12589 (2009)

\bibitem{Sardanyes2008}
Sardany\'es, J., Elena, S. F., Sol\'e, R. V. Simple quasispecies models for the survival-of-the-flattest effect: The role of space. J. theor. Biol. 250(3), 560-568 (2008)

\bibitem{Seger1992}
Seger, J. Evolution of exploiter-victim relationships. In \emph{Natural enemies: the population biology of predators, parasites and diseases} (ed. M. J. Crawley), pp. 3-25. Oxford, UK: Blackwell Scientific (1992)

\bibitem{Sole2006}
Sol\'e, R. V., Sardany\'es, J., D\'iez, J., Mas, A. Information catastrophe in RNA viruses through replication thresholds. J. theor. Biol. 240(3), 353-359 (2006)

\bibitem{Sole2003}
Sol\'e, R. V. Phase transitions in unstable cancer cell populations. Europ. Phys. Journal 35(1), 117-124 (2003)

\bibitem {Sole1999}
Sol\'e, R. V., Ferrer, R., Gonz\'alez-Garc\'ia, Quer, J., Domingo, E. Red Queen dynamics, competition and critical points in a model of RNA virus quasispecies. J. theor. Biol. 198, 47-59 (1999)

\bibitem{SoleBascompte-SO} 
Sol\'e, R. V., Bascopmte, J. \emph{Self-organization in Complex Ecosystems}. Princeton University Press: New Jersey (2006)

\bibitem{Sole1996}
Sol\'e, R. V. On macroevolution, extinctions and critical phenomena. Complexity 1(4), 40-46 (1996)

\bibitem{Sole1996a}
Sol\'e, R. V., Bascompte, J., Manrubia, S. C. Extinction: bad genes or weak chaos? Proc. Roy. Soc. B 263, 161-168 (1996)

\bibitem{Sole1996b}
Sol\'e, R. V., Manrubia, S. C. Extinction and self-organized criticality in a model of large-scale evolution. Phys. Rev. E 51, 6250-6253 (1996)

\bibitem{Sole1997}
Sol\'e, R. V., Manrubia, S. C. Criticality and unpredictability in macroevolution. Phys. Rev. E 55, 4500-4508 (1997)

\bibitem{Sole2002}
Sol\'e, R. V., Montoya, J., Erwin, D. H. Recovery after mass extinction: evolutionary assembly in large-scale biosphere dynamics. Phil. Trans. Roy. Soc. B-Biol. Sci. 357, 697-707 (2002)

\bibitem{Sole2010}
Sol\'e, R. V., Salda\~na, J., Montoya, J. M., Erwin, D. H. Simple model of recovery dynamics after mass extinction. J. Theor. Biol. 267, 193-200 (2010)

\bibitem{Sole1998}
Sol\'e, R. V., Manrubia, S. C., Mercader, J. P., Benton, M., Bak, P. Long-range correlations in the fossil record and the fractal nature of macroevolution. Adv. Compl. Syst. 1, 255-266 (1998)

\bibitem{Stenseth1984}
Stenseth, N. C., Maynard Smith, J. Coevolution in ecosystems: Red Queen evolution or stasis? Evolution 38, 870-880 (1984)

\bibitem{Tegmark1996}
Tegmark, M. An icosahedron-based method for pixelizing the celestial sphere. The Astrophys. J. 470, L81-L84 (1996)

\bibitem{Thompson1982}
Thompson, J. N. \emph{Interaction and coevolution}. Wiley: New York (1982)

\bibitem{Thompson1989}
Thompson, J. N. Concepts of coevolution. Trends Ecol. Evol. 4, 179-183 (1989)

\bibitem{Thompson1990}
Thompson, JN. in \emph{Pests, Pathogens and plant communities} (eds Burdon, J. J. \& Leather, S. R.) 249-271. Blackwell: Oxford (1990)

\bibitem{Thompson1992}
Thompson, J. N., Burdon, J. J. Gene-for-gene coevolution between plants and parasites. Nature 360, 121-125 (1992)

\bibitem{VanValen1976}
Van Valen, L. Energy and evolution. Evol. Theory 1, 179-229 (1976)

\bibitem{VanValen1980}
Van Valen, L. Evolution as a zero-sum game for energy. Evol. Theory 4, 129-142 (1980)

\bibitem{vanValen1973}
Van Valen, L. A new evolutionary law. Evol. Theory 1, 1-30 (1973)

\bibitem{Vidal1984}
Vidal, C., Pacault, A., Eds. \emph{Non-Equilibrium Dynamics in Chemical Systems}. Springer Series in Synergetics; Springer-Verlag: New York (1984)

\bibitem{Weltz2006}
Weltz, J.S., Levin, S. Size and scaling of predator-prey dynamics. Ecol. Lett. 9, 548-557 (2006)

\bibitem{Wright1931}
Wright, S. Evolution in Mendelian populations. Genetics 16, 97 (1931)

\bibitem{Wright1932}
Wright, S. The roles of mutation, inbreeding, crossbreeding and selection in evolution. Proceedings of the Sixth International Congress on Genetics 1, 356 (1932)

\bibitem{Yip2008}
Yip, K.Y., Patel, P., Kim, P.M., Engelman, D.M., McDermott, D., Gerstein, M. An integrated system for studying residue coevolution in proteins. Bioinformatics 24, 290-292 (2008)

\end{thebibliography}
%

\end{document}